\documentstyle[11pt,aaspp4,flushrt,tighten,psfig]{article}  

\begin{document}

\title{The Evolution of X--ray Clusters and the Entropy of the
Intra--Cluster Medium}

\author{Paolo Tozzi\altaffilmark{1}}

\and
\author{Colin Norman}
\affil{Department of Physics and Astronomy, Johns Hopkins University,
Baltimore, MD 21218.} 
\altaffiltext{1}{Osservatorio Astronomico di Trieste, via Tiepolo 11, I-34131
Trieste, Italy}

\begin{abstract}
The thermodynamics of the diffuse, X-ray emitting gas in clusters of
galaxies is determined by gravitational processes associated with
infalling gas, shock heating and adiabatic compression, and
non--gravitational processes such as heating by SNe, stellar winds,
activity in central galactic nuclei, and radiative cooling.  The
effect of gravitational processes on the thermodynamics of the Intra
Cluster Medium (ICM) can be expressed in terms of the ICM entropy.
The entropy is a convenient variable as long as cooling is negligible,
since it remains constant during the phase of adiabatic compression
during accretion into the potential well, and it shows a single
step-like increase during shock heating.  Observations indicate that
non--gravitational processes also play a key role in determining the
distribution of entropy in the ICM.  In particular an entropy excess
with respect to that produced by purely gravitational processes has
been recently detected in the centers of low temperature systems. This
type of entropy excess is believed to be responsible for many other
properties of local X--ray clusters, including the $L$--$T$ relation
and the flat density cores in clusters and groups.

In this paper we assume that the entropy excess is present in the
Intergalactic Medium (IGM) baryons before the gas is accreted by the
dark matter halos and reaches high densities.  We use a generalized
spherical model to compute the X--ray properties of groups and
clusters for a range of initial entropy levels in the IGM and for a
range of mass scales, cosmic epochs and background cosmologies.  In
particular, we follow the formation of adiabatic cores during the
first stages of the gravitational collapse, and the subsequent
evolution of the central entropy due to radiative energy loss. The
model predicts the statistical properties of the cluster population at
a given epoch, and also allows study of the evolution of single X--ray
halos as a function of their age.

We find that the statistical properties of the X--ray clusters
strongly depend on the value of the initial background entropy.
Assuming a constant, uniform value for the background entropy, the
present--day X--ray data are well fitted for the following range of
values of the adiabatic constant $K_*\equiv k_BT/\mu m_p \rho^{2/3} =
(0.4\pm 0.1) \times 10^{34}$ erg cm$^2$ g$^{-5/3}$ for clusters with
average temperatures $kT>2$ keV; $K_* = (0.2\pm 0.1) \times 10^{34}$
erg cm$^2$ g$^{-5/3}$ for groups and clusters with average
temperatures $k_BT<2$ keV.  These values correspond to different
excess energy per particle of $k_BT \geq 0.1 (K_*/0.4\times 10^{34}) $
keV.  The dependence of $K_*$ on the mass scale can be well reproduced
by an epoch dependent external entropy: the relation $K_* =
0.8(1+z)^{-1}\times 10^{34}$ erg cm$^2$ g$^{-5/3}$ fits the data over
the whole temperature range.  The model can be extended to include
internal heating, but in this case the energy budget required to fit
the X--ray properties would be much higher.  Observations of both
local and distant clusters can be used to trace the distribution and
the evolution of the entropy in the cosmic baryons, and to constrain
the typical epoch and the source of the heating processes.  The X--ray
satellites Chandra and XMM can add to our knowledge of the history of
the cosmic baryons, already derived from the high redshift, low
density gas observed in the QSO absorption-line clouds, by imaging the
hot, higher density plasma observed in groups and clusters of
galaxies.
\end{abstract}


\section{INTRODUCTION}

Clusters of galaxies are the largest virialized objects in the
Universe, and are usually considered a canonical data set for testing
cosmology. They are the largest collections of diffuse, highly
ionized baryons that are directly observable in X--rays mostly through
thermal bremsstrahlung emission.  The strong dependence of X--ray
emission on density $L\propto \rho^2$ allows one to select
clusters and define complete samples much better than in the optical
band.

X-ray observations of cluster number counts, luminosity functions and
temperature distributions indicate little apparent evolution in
clusters back to redshifts as high as $\sim 0.7$ (e.g., Henry 1997,
2000; Rosati et al. 1998; Schindler 1999), with the exception of very
high luminosity objects or very high redshifts (Gioia et al. 1990;
Rosati et al. 2000).  This set of results provides one of the
strongest challenges to high--density cosmological models in which
cluster evolution is expected to be detectable even at redshifts as
low as $z\simeq 0.3$. However, these tests are highly dependent on the
thermodynamic evolution of the ICM (e.g. see Borgani et al. 1999 and
references therein; Bower 1997).  The best--fit cosmological
parameters are degenerate with the phenomenological parameters used to
describe the evolutionary properties of the ICM.  In fact, the diffuse
baryons in clusters do not simply follow the dark matter, as would be
the case if they were driven only by gravity as in self--similar
models (Kaiser 1986).  Significant efforts have been devoted recently
to building a physical model for the ICM including an energy scale at
which baryons and dark matter effectively decouple and the
self--similarity is broken.

The presence of a minimum entropy in the pre-collapse IGM has been
advocated for some time as a way to naturally break the self-similar
behaviour (Kaiser 1991, Evrard \& Henry 1991).  Such an extra entropy
is the key ingredient in reproducing the observed
luminosity--temperature relation $L\propto T^{n}$ with $n\simeq 3$
(David et al. 1993, Mushotzky \& Scharf 1997, Allen \& Fabian 1998;
Arnaud \& Evrard 1999; Markevitch 1998), which is at variance with the
self--similar prediction $L \propto T^2$.  Such an entropy minimum
bends the relation from self--similar $L\propto T^2$ behaviour at very
large scales ($\sim 10^{15} M_\odot$) towards a steeper slope on the
scale of groups ($\sim 10^{13}- 10^{14}M_\odot$) which is actually
observed (Ponman et al. 1996; Helsdon \& Ponman 2000).  The average
$L\propto T^3$ relationship is essentially produced by the flattening
of the density distribution in the cores of the X--ray halos; such
cores grow larger as the mass scale decreases, and the luminosity
steepens further on the scale of groups, where the gas is only
adiabatically compressed (see Balogh, Babul \& Patton 1999; Cavaliere,
Menci \& Tozzi 1997, hereafter CMT97; Cavaliere, Menci \& Tozzi 1999).

The picture has been reinforced by the net change observed in the
chemical properties and the spatial distribution of the ICM on the
scale of groups, below the observed temperature of 1 keV (Renzini
1997, 1999) where the effects of the entropy excess are expected to be
strongest.  Another piece of evidence can be obtained from the
observed mass--temperature relation (see Horner, Mushotzky \& Scharf
1999).  Recently, an excess of entropy (with respect to the self
similar scaling) has been directly detected in the central regions of
small clusters with temperatures between $1$ and $3$ keV (Ponman,
Cannon \& Navarro 1999, hereafter PCN; see also Lloyd--Davies, Ponman
\& Cannon 2000), pointing to the role of the entropy as the key
ingredient determining the different properties of clusters and
groups.

Independent hints come from the extragalactic X--ray background:
without a substantial entropy injection at early epochs, its level and
correlation function would exceed the observed limits, due to the
widespread cooling phenomena that would radiate most of the
gravitational energy of the collapsing baryons in the soft X--ray band
(Pen 1999; Wu, Fabian \& Nulsen 1999).

However, even if there are many hints pointing towards a comprehensive
picture, there is a large uncertainty on the amount of {\sl
extra--energy} that effectively generates the entropy excess.  It can
be shown that it is the final entropy distribution that determines
both the spatial distribution of the ICM and its evolutionary
properties, irrespective of the total energy released in the past.  A
given entropy level can be reached through different thermodynamic
histories, so that it is not possible to relate the ICM properties
directly to a given energy excess without knowing the detailed physics
of the heating processes.  As we will show in this paper, the first
question to answer is not: {\sl how much energy has been released in
the ICM}? but rather: {\sl what is the sequence of adiabats through
which the baryons evolve}?

It is difficult to predict {\sl a priori} the entropy excess of the
cosmic baryons, since most of the processes regulating nuclear
activity, star and galaxy formation, and the transfer of energy to the
surrounding baryons, are out of reach of present--day techniques.
Thus, at present there is no general consensus on the production
mechanism of such extra--entropy.  For example, it is not clear
whether the entropy minimum has been established in the IGM before it
has been accreted --the external scenario, or in the high density ICM
after accretion--the internal scenario.  A different energy budget is
required in the two different scenarios: a few tenths of a keV per
particle are needed if the entropy is generated early enough to keep
the baryons on a high adiabat, which prevents them from reaching high
densities and cooling massively; much higher energy excess ($> 1$ keV
per particle) is required if the entropy is generated later, when the
cooling process is eventually already widespread and most of the gas
is already at high densities (Tozzi, Scharf \& Norman 2000, hereafter
TSN00).

The external scenario, which we will assume as a reference model, is
provided by a ubiquitous entropy {\sl floor} in the diffuse gas, which
is entirely due to non--gravitational processes and is assumed to be
in place before the onset of gravitational collapse of massive halos.
The initial extra entropy is ineffective in large mass systems, where
most of the entropy is due to strong shocks, but is more important in
smaller mass systems, where the entropy production via shocks is
strongly reduced. Eventually a large part of the baryons are merely
adiabatically compressed and retain full memory of the initial entropy
level.  The non--gravitational origin of the excess entropy is
crucial, since its level is independent of the mass scale and it
breaks the self--similarity, while gravitational processes always
scale self--similarly with mass.

We present a detailed model to relate the thermodynamic properties of
the ICM in groups and clusters of galaxies to an initial entropy
excess in the IGM, taking into account the transition between the
adiabatic and the shock regime in the growth of X--ray emitting
halos. The effect of radiative cooling is also included.  We show
that, despite the many complexities involved, the entropy is always a
convenient synthetic quantity to describe the thermodynamic history of
the cosmic baryons at least on the scale of groups and clusters.  In
particular, we show that in many circumstances the entropy track of a
shell of baryons being accreted onto dark matter halos goes through
three major regimes: (1) adiabatic compression, during which both
heating and cooling are negligible and the entropy is constant; (2)
step--like discontinuities due to gravitationally induced shocks; and
(3) slow decrease when cooling becomes efficient for baryons in the
inner regions of large halos.  The entropy jump, the onset of cooling,
and the final spatial distribution of the ICM, depend on the initial
entropy.  Such an external, initial entropy level can be reconstructed
from the observation of a large number of distant clusters, or from
the spatially and spectrally resolved profiles of nearby halos (see
TSN00).  Even if the knowledge of the entropy does not resolve the
details of the underlying heating history and determine unambiguously
the energy budget, the combination of data in the X-ray band with data
in the optical and infrared bands can help to identify the major
source of heating. In principle, this allows a detailed reconstruction
of the energetic processes that affect the cosmic baryons over a wide
range of scales and cosmic epochs.

The paper is organized as follows.  In \S 2 we establish a
one--to--one correspondence between the entropy level and the
distribution of the ICM in halos in equilibrium.  In \S 3 we present a
generalized spherical infall model to follow the entropy track of each
shell.  In \S 4 we derive the average density and temperature profiles
and the related global properties such as luminosity, emission
weighted temperature and core radius, as a function of mass scale,
cosmology, epoch and dark matter profile.  In \S 5 we widen the
parameter space, and investigate a time--dependent background entropy
to show how the evolution in the entropy reflects in the X--ray
properties of clusters of galaxies.  In \S 6 we discuss the limitation
of the present approach.  Finally, our conclusions and future
perspectives are presented in \S 7.

\section{ICM THERMODYNAMICS: ENTROPY}

The position, density and temperature of each shell in hydrostatic
equilibrium in a given dark matter halo (whose average properties are
determined by its total virialized mass $M_0$ at the epoch of
observation $z_0$), can be unambiguously recovered once the final
entropy profile is known.  Assuming a spherical mass distribution, the
equation of hydrostatic equilibrium for diffuse baryons in the
potential well is:
\begin{equation}
{{1}\over{\rho}} {{dp}\over{dx}} = 
-C{{m(<x)}\over{x^2}}\, ,
\label{eq}
\end{equation}
where the radius $x$, the pressure $p$ and the density $\rho$ refer to
the baryons and are normalized to the respective values at the last
accreted shell at $z=z_0$, while $m$ is the total mass profile
normalized to the total virialized mass.  Explicitly, $x\equiv R/R_s$,
$p\equiv P/P_s$, $\rho = \rho_B/\rho_s$, and $m(<x)\equiv M(<x)/M_0$.
Since dark matter and baryons are distributed differently, we write
$M(<x) = M_{DM}(<x) + M_B(<x) $.  The constant is $C = -GM_0 \mu
m_p/R_s k_BT_s$, where $m_p$ is the proton mass, $G$ is the
gravitational constant, $k_B$ is the Boltzmann constant and $\mu$ is
the molecular weight of the plasma (we will assume $\mu \simeq 0.59$
for a primordial IGM).  $T_s$ is the temperature of the last accreted
baryonic shell.  In the following we will refer to the values of the
last accreted shell as the shock value, even in the limit of a
vanishingly small shock.  We assume that hydrodynamic equilibrium is
instantaneously established after each accretion event.

We define the adiabat $K\equiv k_B T/\mu m_p \rho_B^{\gamma-1}$
(following the notation of Balogh, Babul \& Patton 1999), where
$S\propto {\rm ln} (K)$ is the entropy and $\gamma$ is the microscopic
adiabatic index which is $\gamma = 5/3$ for a monoatomic gas. Using
the perfect gas equation, we can write the density in terms of
pressure and entropy normalized to the value at the last accreted
shell, with each shell scaled to the corresponding adiabat: $\rho=
p^{1/\gamma}\, k^{-1/\gamma}$, where $k\equiv K(x)/K_s$.  Substituting
in Equation (\ref{eq}), the equilibrium pressure profile is re-written
as:
\begin{equation}
{{dp}\over{dx}} = 
-C p^{1/\gamma}\, k^{-1/\gamma}  {{m(<x)}\over{x^2}}\, .
\label{model}
\end{equation}
The above expression allows us to calculate the thermodynamic
properties of a hydrostatic distribution of gas when the adiabat
profile $K(x)$ is known.  The main difference from the usual solutions
of the hydrostatic equilibrium equation is that there is no need to
assume a polytropic index, since each shell already sits on its
adiabat which is determined by its previous history, and the
correspondence between density and temperature is unambiguous.

The problem reduces to finding the proper adiabat of each infalling
shell, or the entropy as a function of the accreted baryons, since the
baryonic mass included in a given shell is constant with time.  This
procedure is convenient when applied to clusters of galaxies, because
the entropy is conserved for the majority of the time.  In fact, the
dynamic history of a shell of gas can be described in three steps: 1)
adiabatic compression during the infall; 2) shock heating at the
accretion; 3) compression within the potential well due to further
growth of the halo. The entropy is therefore constant during the first
and third phase, and the jump at the shock is the most important
feature needed to reconstruct the final profile. Cooling introduces
further complexity, because for the inner, higher density shells, the
radiative loss becomes important, changing substantially the final
adiabats with respect to the initial value.  However, as we will see
later, the cooling can be included in the above picture, as long as
the initial adiabat is not too low.


To begin with, we focus on the most important event in the entropy
history of each shell: the accretion epoch.  To calculate the value of
$K$ immediately after the accretion shock, we need to estimate both the
density and the temperature of each shell after shock heating
eventually raised the adiabat from the external value to the
post--accretion value $K_i(x)$.  If a shock does not occur, the baryons
are only adiabatically compressed and are accreted with the same
adiabat.  To determine whether a shell is shocked or not during
accretion, we build a spherical infall model for the baryons,
generalized for different cosmologies and epochs.

\section{A GENERALIZED SPHERICAL MODEL}

In the framework of the hierarchical clustering scenario, the baryons
are accreted along with the dark matter during the process of
gravitational collapse.  An expanding accretion shock at the interface
of the inner hydrostatic gas with a cooler, adiabatically--compressed,
external medium, located approximately at the virial radius of the
cluster, is a longstanding prediction from such
gravitationally--driven models (see the 1D models of Bertschinger
1985, Ryu \& Kang 1997, Knight \& Ponman 1997, Takizawa \& Mineshige
1998, and the 3D numerical simulations of Evrard 1990, Roettiger et
al. 1993, Metzler \& Evrard 1994, Bryan \& Norman 1998, Abadi, Bower
\& Navarro 2000).  Due to the growth of the total virialized mass, the
baryons accreted later experience larger shocks, and the resulting
entropy profile is always growing outwards.  Such
gravitationally--driven models predict X-ray properties which scale
self--similarly with mass and fail to reproduce the X-ray observations
of clusters.

A non--negligible value of the background entropy is needed in order
to break the self--similarity. In fact, an initial adiabat will
prevent shocks occurring below a given mass scale.  We now discuss the
external scenario in which an initial adiabat $K_*$ is imprinted on
all the diffuse IGM at some epoch prior to the formation of the dark
matter potential wells.  We refer to $K_*$ as to the background entropy
established in the IGM by non--gravitational processes before the
baryons are accreted.

\subsection{Accretion and Shock Conditions}

The most prominent feature of the entropy history of each shell is the
discontinuity at the accretion shock.  To calculate the discontinuity
we need to know the pre--shock density and the temperature that the
infalling gas reaches moving along the initial adiabat $K_*$ {\sl
before} accretion.  Then we calculate the postshock temperature and 
density using mass, momentum and energy conservation, in the limit of
complete thermalization of the kinetic energy of the gas.

The first important quantity is the infall velocity $v_i$. The
dependence of $v_i$ on the total mass enclosed by the shell can be
written as:
\begin{equation} 
{{ v_i^2}\over 2} = {{ v_{ff}^2}\over 2} + \Delta W  - {{ c_{s}^2}\over
{\gamma -1}} + {{ c_{s}^2}\over {\gamma -1}} \Big({{ \rho_{ta}}\over
{\rho_e}}\Big)^{\gamma -1}\, , \label{vi} 
\label{v_i}
\end{equation}
where $\rho_{ta}$ is the density at turnaround, $\rho_e$ is the gas
external density, $c_s=\sqrt{\gamma K_*\rho_e^{\gamma -1}}$ is the
sound speed (hereafter $\gamma = 5/3$), both calculated at the
accretion radius $R_s$, and $v_{ff}$ is the free--fall velocity of a
particle containing always the same amount of mass during the infall.
Equation \ref{v_i} is a generalized version of the Bernoulli equation
for an adiabatic, spherically simmetrical accretion (Bondi 1952).  The
last quantity can be written as:
\begin{equation}
{{v_{ff}^2}\over 2}\equiv
{{GM}\over{R_s}}-{{GM}\over{R_{ta}}}\, ,
\label{approx}
\end{equation}
where $M$ is the total mass initially included by the baryonic shell.
The term $\Delta W$ is the contribution added to $v_{ff}^2/2$ to
obtain the total work done by the gravitational potential on the
baryonic shell, from the turnaround radius, $R_{ta}$, to the accretion
radius $R_s$, including the effect of the time--varying enclosed mass.
To evaluate this term it is strictly necessary to solve the trajectory
of each baryonic shell.  However we can make the simplifying
assumption that the amount of dark matter enclosed by each shell, is a
monotonically growing function of time, from the mass enclosed at turn
around, to the final mass enclosed at the shock radius.  The term
$\Delta W$ is estimated in \S A, and the uncertainty on it turns out
to be approximately $10$--$30$ \%.  We show later that this error is
not important in determining the transition scale between the shock
and the adiabatic regime.

The other two terms proportional to $c_s^2$ describe the energy needed
to compress the gas.  In fact, due to the non--negligible value of
$K_*$ in the infalling IGM, part of the gravitational energy goes into
internal energy in an amount proportional to the square of the sound
speed in the external IGM at the epoch of accretion, so that in
general $v_i < v_{ff}$.  The compression term carries an increasing
fraction of the potential energy when the mass of the system is lower,
or, since the sound speed is proportional to $K_*^{1/2}$, when the
entropy is higher.  The fourth term on the right hand side of Equation
(\ref{v_i}) results from the initial condition $v_i=0$ for a gas shell
at the turnaround radius, when the gas had a density $\rho_{ta}$ and
it is assumed to be at the same contrast of the dark matter.  The
epoch of turnaround is assumed to be half of the infall epoch.

Of course to solve Equation (\ref{v_i}) we need to evaluate $\rho_e$.
To do this, we first note that the knowledge of both the external
density and the infall velocity gives the net infall accretion rate of
baryonic matter through the surface defined by the shock radius.
Then, we make the assumption that {\sl the growth rate of the total
virialized mass $\dot M$ is proportional to the growth rate of the
thermalized baryonic mass $\dot M_B$}.  Here $\dot M$ is the average
total mass accretion rate as predicted in the hierarchical clustering
scenario.  This means that all the baryons, that initially
were in the same lagrangian volume of the mass that is currently
virialized, have been accreted.  The proportionality constant is
simply the average mass fraction of baryons in diffuse form $f_B$, so
that at each epoch the fraction of accreted baryons (with respect to
the total baryons accreted at $z=z_0$), is equal to the fraction of
the accreted matter to the total virialized mass at the same final
epoch. This does not imply that the baryons are in the same
volume; they are distributed in a volume typically larger than that of
the accreted dark matter.  This occurs especially in the adiabatic
regime, when the baryons have too high a temperature to sink into the
potential well and thus the accretion radius is significantly larger
than the virial one.  The constraint on the mass accretion rate
translates into the relation:
\begin{equation}
\dot M_B = f_B \dot M = \rho_e 4 \pi R_S^2 
\Big( v_i + {{dR_S}\over{dt}}\Big)\, ,
\label{mdot}
\end{equation}
where $\dot M$ is given for a particular cosmological model (see \S
3.3).  We can derive $\rho_e$ as a function of $v_i$, and then the
external temperature is $k_BT_e=\mu m_p K_*\rho_e^{2/3}$.

The condition $v_i>c_s$ determines if the shell is shocked. In the
frame of the infalling gas the shock expands with a velocity
$v_i+{{dR_S}/{dt}}$.  In the case of a shock, we assume that all the
kinetic energy of the infalling gas is thermalized (i.e., the
post-shock velocity $v_{ps} = 0$ in the rest--frame of the cluster),
and obtain for the postshock temperature (Landau \& Lifshitz 1957;
Cavaliere, Menci \& Tozzi 1998):
\begin{equation}
k_BT_i = {{\mu m_p v_i^2}\over 3}\Big[ {{(1+\sqrt{1+\epsilon})^2}\over 4}
+ {7\over{10}}\epsilon -{{3}\over {20}}{{\epsilon^2}\over{(1+\sqrt{1+
\epsilon})^2}}\Big]\, ,
\label{ti}
\end{equation}
where $\epsilon\equiv 15 k_BT_e/4 \mu m_p v_i^2$.  

The postshock density is then $\rho_i =g\, \rho_e$, where $g$ is the
shock compression factor which depends on the
postshock temperature, $T_i$, and the external temperature, $T_e$, and
is given by (see CMT97):
\begin{equation}
g = 2\,\Big(1-{T_e\over T_i}\Big)+\Big[4\, 
\Big(1-{T_e\over T_i}\Big)^2 + {T_e\over T_i}\Big]^{1/2}\, .
\label{compression}
\end{equation}

If the gas is shocked, we calculate the new adiabat $K_i=k_BT_i/\mu
m_p \rho_i^{2/3}$ of the baryonic shell after accretion.  If the
infalling velocity is smaller than the sound speed in the external IGM
and the shock does not occur, the gas is accreted adiabatically, and
therefore the post--accretion adiabat is the inital one $K_i=K_*$,
which is all we need to solve for the final equilibrium.

\begin{figure}
\centerline{\psfig{figure=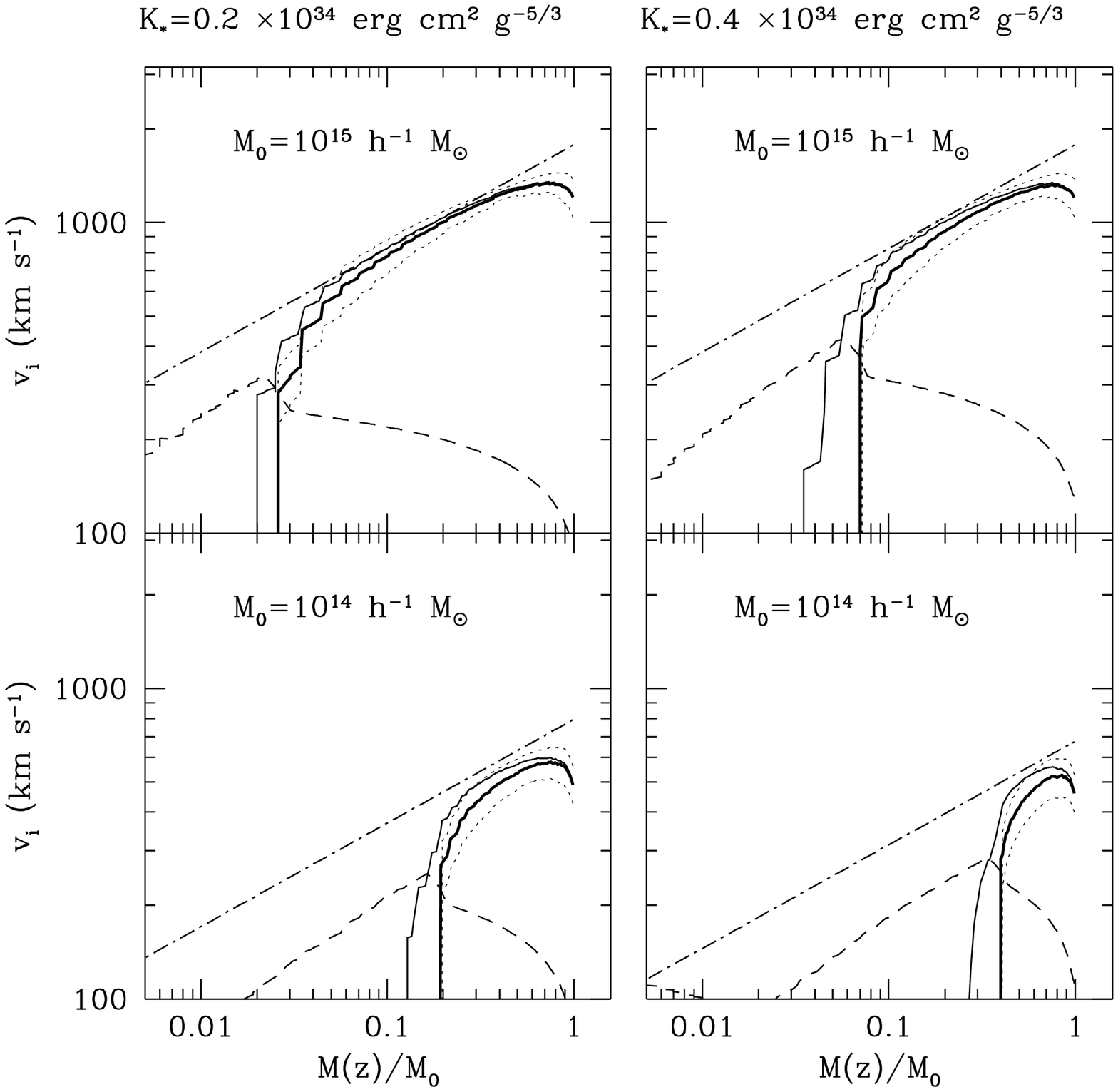,height=15truecm,width=15truecm}}
\caption{Thick solid line shows the infall velocity $v_i$
computed at the shock radius of each baryonic shell as a function of
the virialized mass (normalized to the final value).  The
uncertainties in $v_i$ are shown by dotted lines.  Here we assumed
$K_* = 0.2$ (left panels) and $K_*= 0.4 \times 10^{34}$ erg cm$^{2}$
g$^{-5/3}$ (right panels) in a low density ($\Omega_0=0.3$) flat
cosmology (see Table 1), for a final mass at $z=0$ of $10^{15}\,
h^{-1} M_\odot$ and $10^{14}\, h^{-1} M_\odot$.  The thin solid
line shows the free--fall velocity at the position of the shock, while
the dashed line shows the sound speed $c_s$ computed in the gas
external to the shock.  When $v_i<c_s$ the accretion process is
entirely adiabatic.  The dot--dashed line shows the asymptotic
behaviour $v\propto m^{1/3}$, which is reproduced in the strong shock
regime, when the shock radius is close to the virial radius.
\label{fig1}}
\end{figure}

Thus, using Equations (\ref{v_i}), (\ref{ti}) and (\ref{compression}),
we are able to associate with each shell, including a mass $M_B$ of
baryons, its postshock adiabat $K_i(M_B)$.  For a given object, the
adiabat of the infalling shells initially will be $K_i=K_*$, since for
sufficiently low velocities the shocks are suppressed.  As the total
mass grows, the velocities of the infalling shells rise approximately
as $v_i\propto M^{1/3}$, more rapidly than the sound speed (which in
general decreases with epoch, since $c_s\propto \rho_e^{1/3}\propto
1+z$), and eventually a shock regime begins.  In Figure \ref{fig1} the
transition between the two regimes is shown as a function of the
accreted mass for a given initial adiabat $K_*$.  As it is shown in
the first panel, the maximum uncertainty in the infalling velocity,
$v_i$, grows toward the adiabatic regime, but it does not introduce a
large error in the transition scale, since the infall velocity falls
steeply below $c_s$.  The rapid increase of both the infall and the
free--fall velocity at the transition, occurs because  the
gravitational energy becomes sufficient to overcome the pressure
support, and the accretion radius moves from a relatively distant position
to a position very close to the virial radius.  Clearly, the presence
of a larger $K_*$ further delays the onset of the shock heating
regime, inhibiting adiabatic accretion for the majority of the baryons,
especially in small mass systems.

At this stage, if we neglect further changes in the entropy, the
adiabat in the final position is simply $K(x)=K_i$ and the Equation
(\ref{model}) can be solved easily without any further steps.
However, for the inner shells, radiative cooling becomes important and
the calculation of the final adiabat requires solving Equation
(\ref{model}) at different epochs, as explained in the following
subsection.

\subsection{The Effect of Radiative Cooling}

Each shell of gas is continuously changing its adiabat due to cooling
and heating processes.  In particular, the first baryonic shells that
are accreted drain into the inner, higher density regions of halos as
the total virialized mass grows, and their cooling times become small
enough to start cooling processes.  As a result, the final adiabat of
these baryonic shells will be lower than that at the accretion epoch,
and eventually part of the gas leaves the diffuse, emitting phase and
sinks into the center.

We can model the cooling assuming a homogeneous, single temperature
distribution (Fabian \& Nulsen 1977, Mathews \& Bregman 1978); in this
case the energy equation can be formally written as:
\begin{equation}
{{d}\over{dt}}[{\rm ln} (K)]=-{1\over {\tau_{cool}(K)}}\, ,
\label{ent_evol}
\end{equation}
where the cooling time $\tau_{cool}$ is defined as:
\begin{equation}
\tau_{cool}\equiv {3\over 2} {{k_B T}\over {\Lambda_{net}}}\,
{{\rho_B}\over{\mu m_p}} ,
\label{t_cool}
\end{equation}
and therefore it depends on $K$ through $T$, $\rho_B$ and
$\Lambda_{net}$.  Here $\Lambda_{net}$ is the cooling rate including
free-free and line emission (see Sutherland \& Dopita 1993).

It is well known that cooling is a runaway process, and the solution
of Equation (\ref{ent_evol}) would require the computation of the
equilibrium profile at many different epochs.  Since we still want to
have the benefit of a relatively fast computation, much faster than a
full hydrodynamic simulation, we tackle the problem choosing a medium
resolution in time ($\Delta t \simeq 0.3$ Gyr) and solving Equation
(\ref{ent_evol}) within $\Delta t$ for every shell with an analytic
approximation.  This is possible if we assume that the cooling process
is isobaric within $\Delta t$, in order to express both density and
temperature as a function of the adiabat $K$ only.  The pressure is
updated at each time step, following the new equilibrium
configuration.  An intermediate step is to approximate the cooling
function, $\Lambda_{net}$, with an analytic function of the
temperature.  In this way the change in the adiabat within $\Delta t$
can be derived as the integral of an analytic function, as described
in Appendix \S B.

When the cooling times become very short in the center of the halo,
part of the gas may eventually cool in a single time step $\Delta t$
(i.e., its entropy drops to zero).  In this case, the gas is removed
from the diffuse, emitting phase, and is included in a gravitational
term as if it is all accumulated in the very center.  At this level,
we do not implement more sophisticated multiphase models which can be
important for the detailed emissivity distribution in cooling flows.
However, we can follow the steepening of the baryonic density in the
center as the radiative cooling becomes efficient, and compute the
corresponding amount of baryons which drop out from the diffuse phase.
We stress the fact that we are able to follow the complex cooling
processes with good accuracy by virtue of the initial entropy level.
The background entropy, in fact, delays and possibly inhibits the onset of
strong cooling flows.  Our model breaks down in the limit of small
initial entropy, where the cooling catastrophe occurs.

\begin{figure}
\centerline{\psfig{figure=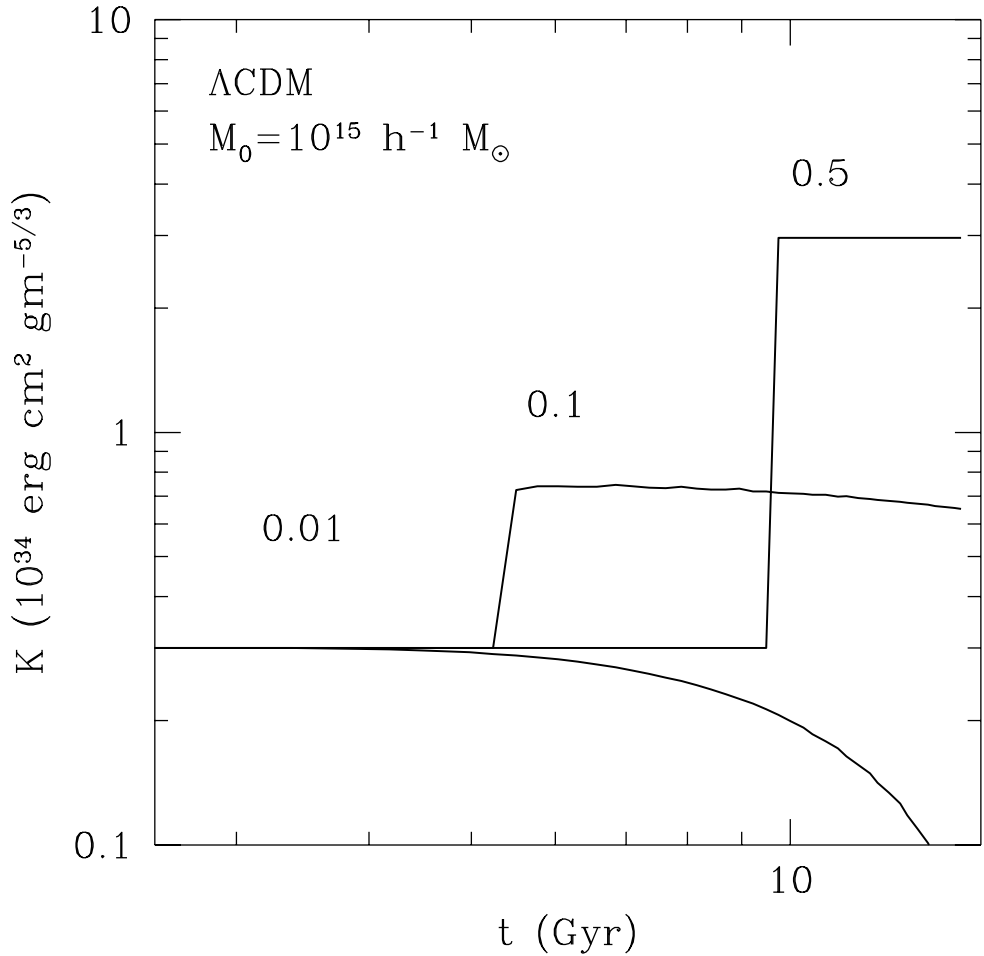,height=15truecm,width=15truecm}}
\caption{The evolution of the adiabat $K$ for three baryonic shells is
shown as a function of cosmic epoch $t$ ($\Lambda$CDM cosmology, for a
final mass of $10^{15} \, h^{-1}\, M_\odot$, $K_* = 0.3 \times
10^{34}$ erg cm$^{2}$ g$^{-5/3}$).  The inner shells contain 1\% of
the total baryons, the second 10\% and the third 50\%.  The sharp
discontinuity, increasing for outer shells, occurs at the shock.
\label{fig2}}
\end{figure}

The evolution of the adiabat as a function of cosmic epoch for some
given shells is plotted in Figure \ref{fig2}.  The outermost shells
are accreted at later epochs.  They are strongly shocked and reach a
high adiabat, and find  equilibrium at large radii and low
densities.  Consequently, the cooling times are always large and the
adiabat $K$ stays almost constant after the accretion.  Conversely,
inner shells are more affected by cooling for two reasons: they reach
much higher densities (being in the central regions), and they have
more time to cool since they are accreted much earlier. Eventually,
the very inner shells reach very low entropy, corresponding to
extremely high densities and very short cooling times, and they
rapidly cool and drop out of the diffuse phase.

\begin{figure}
\centerline{\psfig{figure=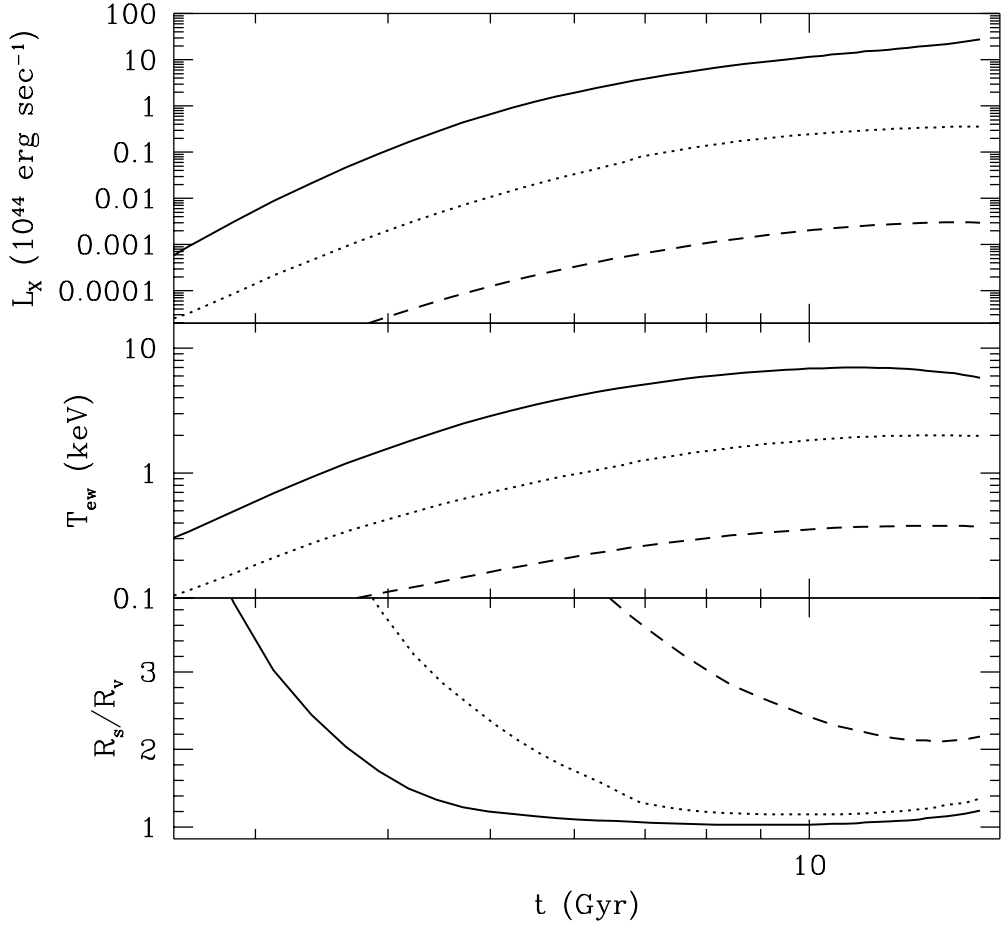,height=15truecm,width=15truecm}}
\caption{The evolution of the total bolometric luminosity $L_X$ and of
the emission weighted temperature $T_{ew}$ is shown as a function of
cosmic epoch for a final mass of $10^{15}h^{-1} M_\odot$ (solid
lines), $10^{14} h^{-1} M_\odot$ (dotted lines) and $10^{13} h^{-1}
M_\odot$ (dashed lines) for a $\Lambda$CDM cosmology, with a constant
initial adiabat $K_* = 0.3 \times 10^{34}$ erg cm$^{2}$ g$^{-5/3}$.
In the third panel the evolution of the shock radius $R_s$, normalized
to the virial radius at each epoch, is shown for the same halos.
\label{fig3}}
\end{figure}

The calculation without the inclusion of cooling would be much
simpler, since the final adiabat would be the accretion value $K_i$
for all the shells, and the hydrostatic equilibrium would be solved
only once (at the final epoch $z_0$).  However, solving the
equilibrium at several epochs allows us to follow the evolution of the
X--ray properties for each (average) dark matter halo.  In Figure
\ref{fig3} the evolution of temperature and luminosity for three
objects of $10^{15} h^{-1}$ (continuous line), $10^{14} h^{-1}$
(dashed line), and $10^{13} h^{-1}\, M_\odot$ (dotted line), is shown
for a constant $K_* =0.3 \times 10^{34}$ erg cm$^2$ g$^{-5/3}$ in a
$\Lambda$CDM cosmology.  In the third panel, the time evolution of the
shock radius is plotted for the same objects.  The shock radius is
normalized to the virial radius at each epoch.  It is possible to see
how the shock radius is close to the virial one for the largest halo
and relaxes in the last few Gyr when the mass accretion slows down and
the external pressure term correspondingly decreases.  The effect is
more pronounced at lower masses, where the internal pressure support
is strong enough to dominate the gravitational potential and the
external pressure term of the infalling gas.

\begin{figure}
\centerline{\psfig{figure=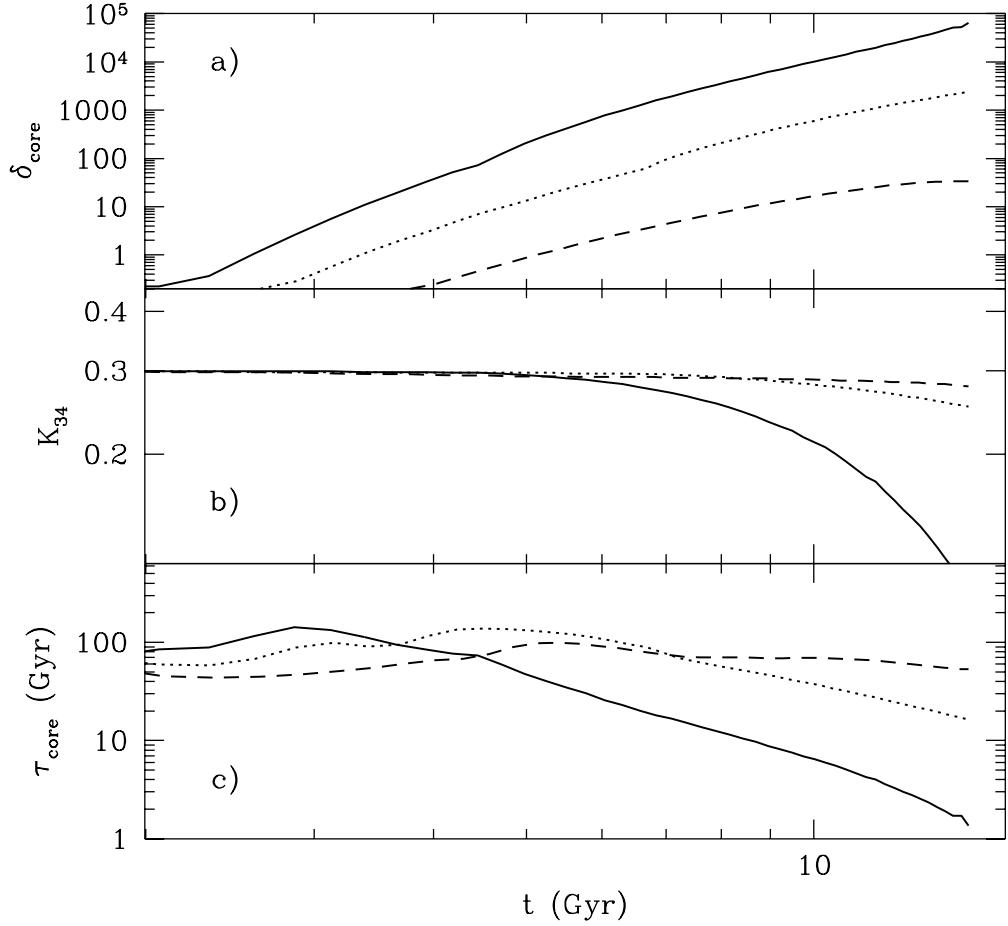,height=15truecm,width=15truecm}}
\caption{The evolution of the properties of the central adiabatic
cores is shown as a function of cosmic epoch for a final mass of
$10^{15}h^{-1} M_\odot$ (solid lines), $10^{14} h^{-1} M_\odot$
(dotted lines) and $10^{13} h^{-1} M_\odot$ (dashed lines) for a
$\Lambda$CDM cosmology, with a constant initial adiabat $K_* = 0.3
\times 10^{34}$ erg cm$^{2}$ g$^{-5/3}$.  Panel a): average
overdensity with respect to the critical value; b) average $K$; c)
average cooling time.
\label{fig4}}
\end{figure}

In Figure \ref{fig4} we plotted, for the same three final masses, some
relevant quantities averaged over the adiabatic cores, defined as
regions including the gas accreted during the adiabatic regime. It is
possible to see how the initial entropy $K_* = 0.3\times 10^{34}$ erg
cm$^2$ g$^{-5/3}$ introduces a large difference in the central core as
a function of mass.  Central densities are much higher for deeper
potential wells. In addition, the baryons in the center of massive
clusters suffer radiative losses and the baryonic cores shrink to
smaller sizes and higher densities.  In the second panel the average
entropy of the inner cores is shown. The decrease due to the radiative
cooling can reduce the initial entropy especially in the most massive
halo. The decrease in the entropy is driven by the decrease in the
average cooling time shown in the third panel.  While the entropy is
decreasing, the internal energy of the gas is still rising due to the
compressional work done by the gravitational potential.  However, the
trend of stronger cooling for larger masses, is reversed in the case
of very small $K_*$.  In fact, as long as there is nothing to prevent
the baryons from cooling, the amount of radiative losses is mainly set
by the age of the halos.

\subsection{Dark Matter Properties}

In this section we briefly review the properties of the dark matter
halos which drive gravitationally the evolution of the diffuse
baryons.  In particular we describe the mass profiles and the mass
accretion rates in the framework of the hierarchical clustering
scenario in universes dominated by cold dark matter (CDM).  However,
the model can be generalized to other cosmologies.

The boundary of a halo is the virial radius, defined as the radius
within which the average overdensity with respect to the critical
density is $\Delta_c$, where $\Delta_c=178$ for $\Omega_0=1$ with a
mild dependence on $\Omega_0$ (see, e.g., Eke et al. 1998).
Analytical studies indicate simple power law profiles for the dark
matter, of the kind $\rho\propto x^{-\xi}$, with $\xi = 9/4$ (Gunn \&
Gott 1972, Bertschinger 1985).  Numerical works show a more complex
behaviour, with a characteristic internal scale radius that depends on
the epoch and on the final mass (Navarro, Frenk and White 1997,
hereafter NFW; Moore et al. 1998).  A very general expression for the
universal profile is:
\begin{equation}
\rho = \rho_{c0} {{\delta_c}\over{(cx/x_v)^{\nu}[1+(cx/x_v)^{\zeta}]^\eta}}\, ,
\label{nfw}
\end{equation}
where $c$ is the mass dependent concentration parameter of the dark
matter, and $\delta_c$ is defined by requiring the average density
within $R_V$ with respect to the critical density to be $\Delta_c$.
Here we used $x_v\equiv R_v/R_s$ to be consistent with Equation
(\ref{model}) where the radius is normalized to $R_s$.  Present
calculations differ mainly in the inner regions, where NFW predict
$\nu =1$, $\zeta =1$ and $\eta =2$, while Moore et al. (1998) have a
steeper inner profile with $\nu =1.5$, $\zeta =1.5$ and $\eta =1$.
From Equation (\ref{nfw}) the mass profiles $m(<r)$ entering Equation
(\ref{model}) follow directly.

We will approximate the concentration parameter with power laws, which
turn out to be good approximations (Navarro et al. 1997). The
expressions used are described in appendix \S C.  In general, the
concentration parameter $c$ depends on the characteristic epoch of
formation of the halo, which in turn depends on cosmology,
perturbation spectrum, $M_0$ and $z_0$ (see NFW). This is because the
dark matter {\sl remembers} the epoch when each shell was accreted,
even if the shell--crossing tends to erase such dynamical memory.  For
example, in a standard CDM universe groups tend to have a larger
concentration ($c\simeq 8$) being formed at higher epochs when the
average density was higher, while clusters, being younger, have a
lower concentration ($c\simeq 6$).  At higher redshifts the
concentration parameters are generally lower, since the difference in
epoch (and thus in typical density) between formation and the
observation epochs $z_0$ is reduced.  These trends will be included in
our calculations.

The accretion processes in groups and clusters show considerable
scatter, as observed in numerical simulations and Monte Carlo
realizations of hierarchical clustering based on the extended Press \&
Schechter formula (hereafter PS, Press \& Schechter 1974, Bond et
al. 1991, Bower 1991, Lacey \& Cole 1993).  However we are interested
in the mass history of {\sl typical} halos, each of them labeled by
the final mass $M_0$ and the final (observation) epoch $z_0$, for a
given cosmology.  The natural way to proceed is to average over many
realizations of the mass history of the {\sl main progenitor}, defined
as the most massive halo participating in every mass accretion event
along the merger tree of a single object.  We run $1000$ Monte Carlo
simulations of the mass history of the main progenitor for different
final masses $M_0$ and different final redshifts $z_0$ ($z_0=0$ and
$z_0=1$) in the two cosmologies discussed below (tCDM and
$\Lambda$CDM).  We find that the average mass growth of the main
progenitor can be approximated within few percent by a parabola in the
${\rm log}(m)$--${\rm log}(1+z)$ space:
\begin{equation}
m(z)=\Big({{1+z}\over{1+z_0}}\Big)^{-[B+A\, {\rm
log}({{1+z}\over{1+z_0}})]}\, ,
\label{mz}
\end{equation}
where $A$ and $B$ depend on cosmology, $M_0$ and $z_0$.  The relation
(\ref{mz}) is used to determine the accretion epoch of each baryonic
shell after Equation (\ref{mdot}), and thus to compute its density,
$\rho_e$, at the accretion shock.

The dispersion in the profiles and in the accretion process is likely
to introduce some dispersion in the resulting X--ray properties, and
is expected to explain partially the intrinsic scatter observed in the
$L$--$T$ relation.  The intrinsic scatter in the emission is certainly
due also to the presence of cooling flows (Allen \& Fabian 1998;
Arnaud \& Evrard 1999), which in turn can be affected by both the
dynamical and the heating history of the gas.  For these reasons, we
focus on typical halos averaging over many different realizations, and
considering the accretion of baryons as a smooth and continuous
process.  These assumptions clearly break down in the case of massive
merger events (see discussion in \S 6).

\section{RESULTS}

Here we present the X--ray properties of groups and clusters of
galaxies in the case of a constant and homogeneous $K_*$ in the
external IGM.  Our reference calculation will be a flat, low density
cold dark matter universe ($\Lambda$CDM), which is currently preferred
on the basis of the measurements of the expansion rate of the universe
from high $z$ SNe (Riess et al. 1998), from the Cosmic Microwave
Background (see Lange et al. 2000; Balbi et al. 2000) and of the
observation of a high baryonic fraction $f_{obs}> 0.06 h^{-1.5}$ in
clusters (see Ettori \& Fabian 1999), which is consistent with
standard nucleosynthesis if $\Omega_0<0.3$ (White et al. 1993). The
baryonic density is assumed to be $\Omega_B = 0.02 h^{-2}$ (Burles \&
Tytler 1999a, 1999b), consistent with the standard primordial
nucleosynthesis scenario.  From the diffuse, X--ray emitting
component, we exclude a fraction which is assumed to be locked in
stars since the beginning, and is choosen to be $20$ \% of the total
baryons in halos (independent of the mass scales and epoch, i.e., we
assume a constant efficiency of star formation).  The fraction of
baryons cooled in the center, instead, is computed at each epoch and
subtracted from the diffuse, X--ray emitting phase.  For comparison,
we will also discuss a tilted cold dark matter universe (tCDM), where
we are forced to adopt a baryonic density $\Omega_B = 0.04 h^{-2}$,
larger than the standard value, in order to be consistent with the
observed baryonic fraction.  The details for the two universes are
shown in Table \ref{tab1}. The values for $A$ and $B$ in the two
universes are determined with a $\chi^2$ fitting of the average mass
histories with the relation (\ref{mz}), and are reported in Table
\ref{tab2}.

\subsection{Density and Temperature Profiles}

\begin{figure}
\centerline{\psfig{figure=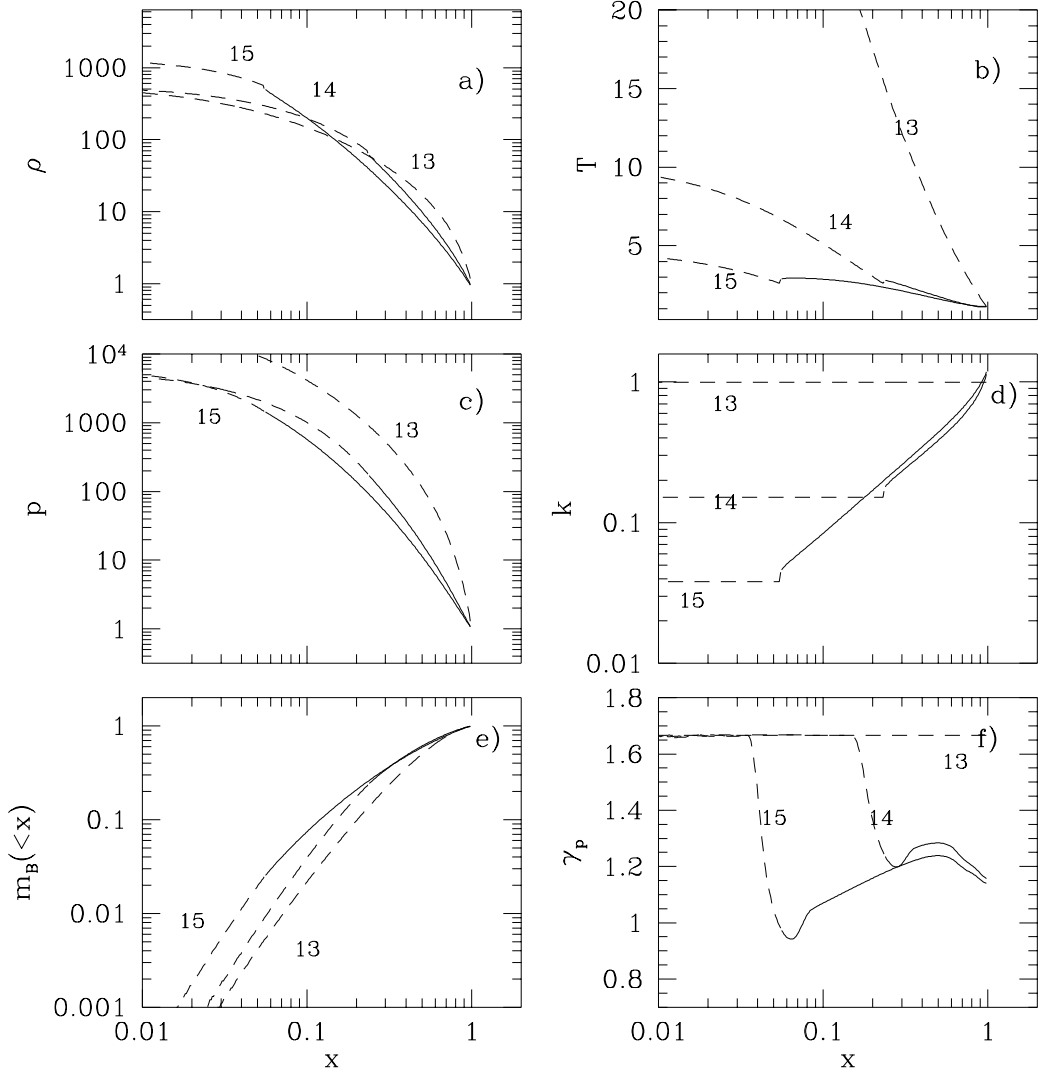,height=15truecm,width=15truecm}}
\caption{Profiles of density, temperature, pressure, entropy,
baryonic mass and polytropic index as a function of the normalized
radius $x\equiv R/R_s$ for clusters of different mass ($10^{15},
10^{14}, 10^{13} h^{-1} M_\odot$ as labeled by the log of the mass) in
$\Lambda$CDM at $z=0$.  Each quantity is normalized with respect to
the corresponding value at the shock, in order to show departures from
self--similarity.  The external, initial adiabat is $K_* =0.3 \times
10^{34}$ erg cm$^2$ g$^{-5/3}$ constant with mass scale and epoch.
The dark matter profiles are from Navarro, Frenk \& White (1997).  No
cooling is assumed.  The dashed lines are for baryons accreted
adiabatically ($K(x)=K_*$), while the solid lines are for the
shocked gas.
\label{fig5}}
\end{figure}

First, we discuss a simple case where the cooling is not included, so
that the final adiabat $K(x)$ is equal to the value at the accretion,
$K_i$.  This case shows the effects of the entropy excess alone
without the intervention of cooling processes.  In Figure \ref{fig5}
we show the resulting profiles for $\Lambda$CDM at redshift $z_0=0$,
for an initial $K_{34} =0.3$, where $K_{34}$ is $K_*$ in units of $
10^{34}$ erg cm$^2$ g$^{-5/3}$.  This value corresponds to a
temperature $k_BT_*\simeq 1.5 \times \, 10^{-2} (1+z)^2$ keV at the
ambient density.  The dark matter is distributed according to the NFW
profile.  Three final masses are shown: $M_0=10^{15}-10^{14}-10^{13}\,
h^{-1} M_\odot$.  The plotted profiles are all normalized at the
corresponding shock values in order to show how the scaling behaviour
departs from self--similarity.  Note, however, that the density and
temperatures values at the shock in physical units are very different
in the three cases.

A characteristic feature is the flat density profile of isentropic gas
in the core, which is relatively larger at smaller masses (dashed
lines in Figure \ref{fig5}, panel a).  Such cores are built in the
initial, high redshift stages of the accretion process, when the
accretion is adiabatic since the infall velocities are small and
shocks do not occur.  This regime is relatively more extended going to
lower masses.  The pressure is more effective in pushing the baryons
over a region larger than that of the dark matter (panel c).  All this
information is synthesized in the entropy profiles: at larger radii
the entropy rises since the outer shells experience stronger shocks
(panel d).  Since the entropy is normalized to the value at the shock
radius, the constant entropy floor in the center appears different at
different masses.  The slope of the entropy profile in the shock
dominated regime is almost independent of the initial value $K_*$,
yielding $d{\rm ln}(K)/d{\rm ln}x \simeq 1.1$; this value is close to
the value $1.3$ expected for the simple case of an isothermal profile
where the entropy is due only to shock heating and $M_B\propto r$.
The sharp knee in the entropy profile is due to the fact that the
transition from adiabatic accretion to strong shocks is very fast, and
the intermediate shock regime virtually does not exist, so that during
the shock regime the entropy is always dominated by shock heating.  In
contrast, in the center, isentropic cores are clearly emerging.  The
ratio of mass accreted adiabatically to the total baryonic mass, is
correspondingly larger at lower mass scales (panel e).

Here we note that a departure from a power--law behaviour for the
entropy profile has been observed in hydrodynamical simulations where
neither radiative cooling nor extra-entropy were included (see Frenk
et al. 1999).  This may suggest that departures from a power--law
behaviour in the entropy profile can also be originated by
asphericity.

The temperature profiles (panel b) do show mild gradients in the
regions where the gas has been shocked (variation less than a factor
of $3$ between $R_s$ and $0.1R_s$), while they show considerable
gradients when the entropy is constant, following $T\propto K_*
\rho^{2/3}$.  Part of the large gradient in the smallest system
corresponds to very low luminosity regions, where the gas is relaxed
due to the very small pressure term.  These regions, and the
corresponding large temperature gradients, have never been observed.
In fact, if we compute the temperature gradient in the inner regions
of halos with $M_0 = 10^{13} h^{-1} M_\odot$, we find an increase of
about 2 within a radius of $0.1 R_S \simeq 100 h^{-1}$ kpc, an effect
hardly visible, e.g., in the data by ROSAT.

A good quantity to characterize the properties of the temperature
profile is the effective polytropic index defined by the relation
$p\propto \rho^{\gamma_p}$.  In general, a family of polytropic
relations can be used to describe the ICM and investigate the energy
budget underlying each polytropic family (Loewenstein 2000).  As
a result of the combined action of shock heating and adiabatic
compression, the index $\gamma_p$ is found to be approximately
$\gamma_p \simeq 0.8 - 1.2$ between the adiabatic core and the shock
radius, roughly consistent with an isothermal temperature profile (in
the Figure we show the value of $\gamma_p$ averaged over $\Delta {\rm
log}(x) = 0.3$).  In the adiabatic cores the polytropic index is
simply $\gamma_p=\gamma=5/3$, since all the gas is on the same
adiabat.

\begin{figure}
\centerline{\psfig{figure=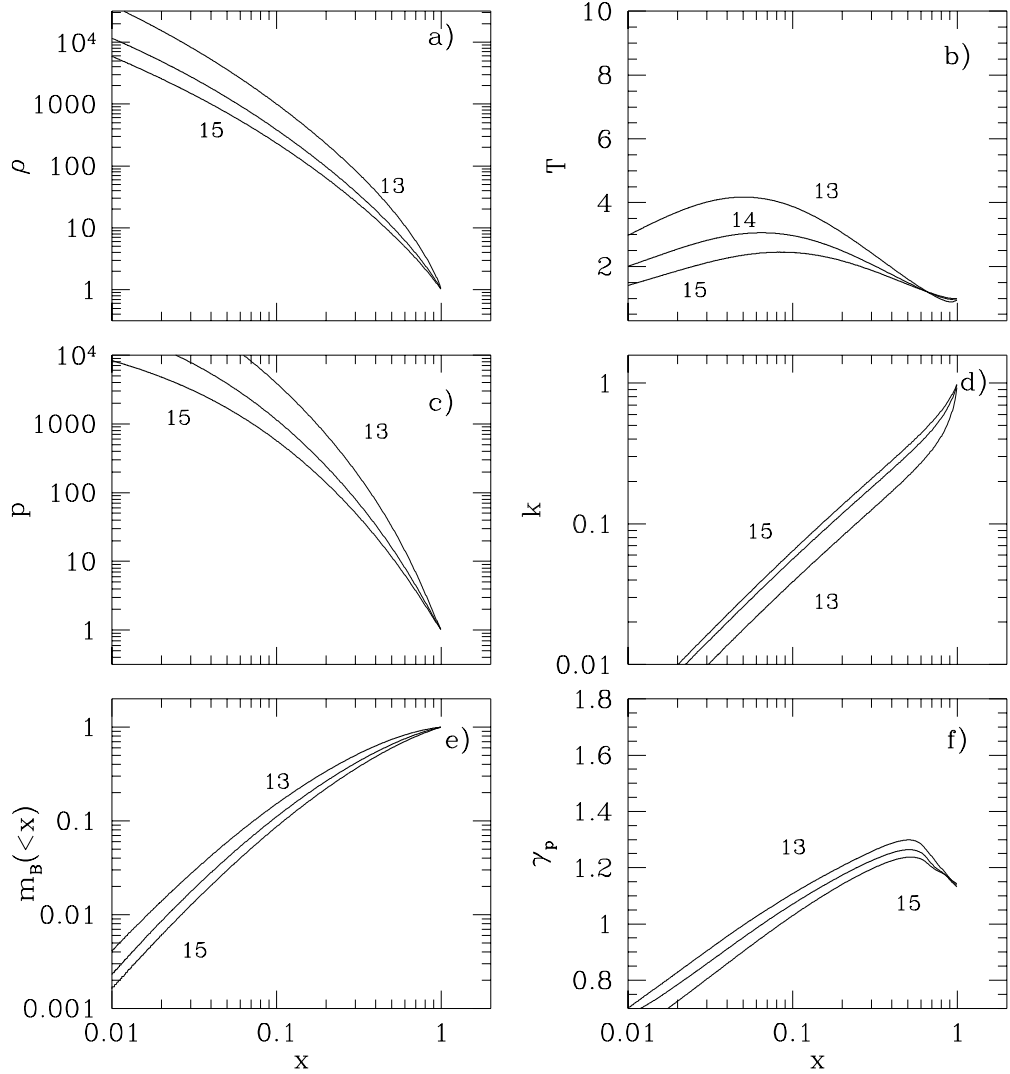,height=15truecm,width=15truecm}}
\caption{The same as in Figure \ref{fig5} for $\Lambda$CDM, but with a
negligible entropy and without the inclusion of cooling (self--similar
case).  All the gas is shocked.  These profiles should be compared
with Figure \ref{fig5}, to point out how the negligible entropy excess
makes the baryons follow the dark matter and produce an opposite
behaviour for which the groups are more centrally concentrated than
clusters, as predicted by NFW.  Note that the entropy profiles  are the
same at all scales.
\label{fig6}}
\end{figure}

To elucidate how the breaking of self--similarity occurs, in Figure
\ref{fig6} we show the same profiles for a negligible value of the
external entropy, but without the inclusion of cooling.  This is what
we call the self--similar case, which is different from the more
realistic case of negligible entropy and the inclusion of cooling,
since cooling also alters the entropy profile, as shown in \S 4.4.  In
the absence of an entropy floor, the profile $K(x)$ always decreases
at smaller radii, and exhibits a power law behaviour without any
particular scale.  The only differences between groups and clusters
are now driven by the dark matter distributions.  Despite the pressure
support, the gas essentially follows the dark matter, and groups
appear more concentrated than clusters, reversing the trend of Figure
\ref{fig5}.

In the $K_* \simeq 0$ case the cooling starts very early and deeply
affects the profiles of massive clusters.  The majority of the
initially diffuse baryons cool in the center of small halos, where,
without an effective background entropy, nothing prevents the baryons
from cooling and the luminosity is dominated by the central regions.
The cooling selectively removes the lower entropy gas in the center of
lower mass objects, helping to create an entropy plateau at the very
center, but with the entropy entirely produced by gravitational
processes.  This mechanism to create an entropy plateau has been
advocated by PCN but a large amount of cooled baryons need to be
accomodated in the center.  The strongest evidence for the presence of
a background entropy at high $z$, is given by the low fraction of
baryons in stars with respect to the total baryons available, which
implies a strong suppression of the cooling processes especially in
low mass halos (see Prunet \& Blanchard 1999).

\begin{figure}
\centerline{\psfig{figure=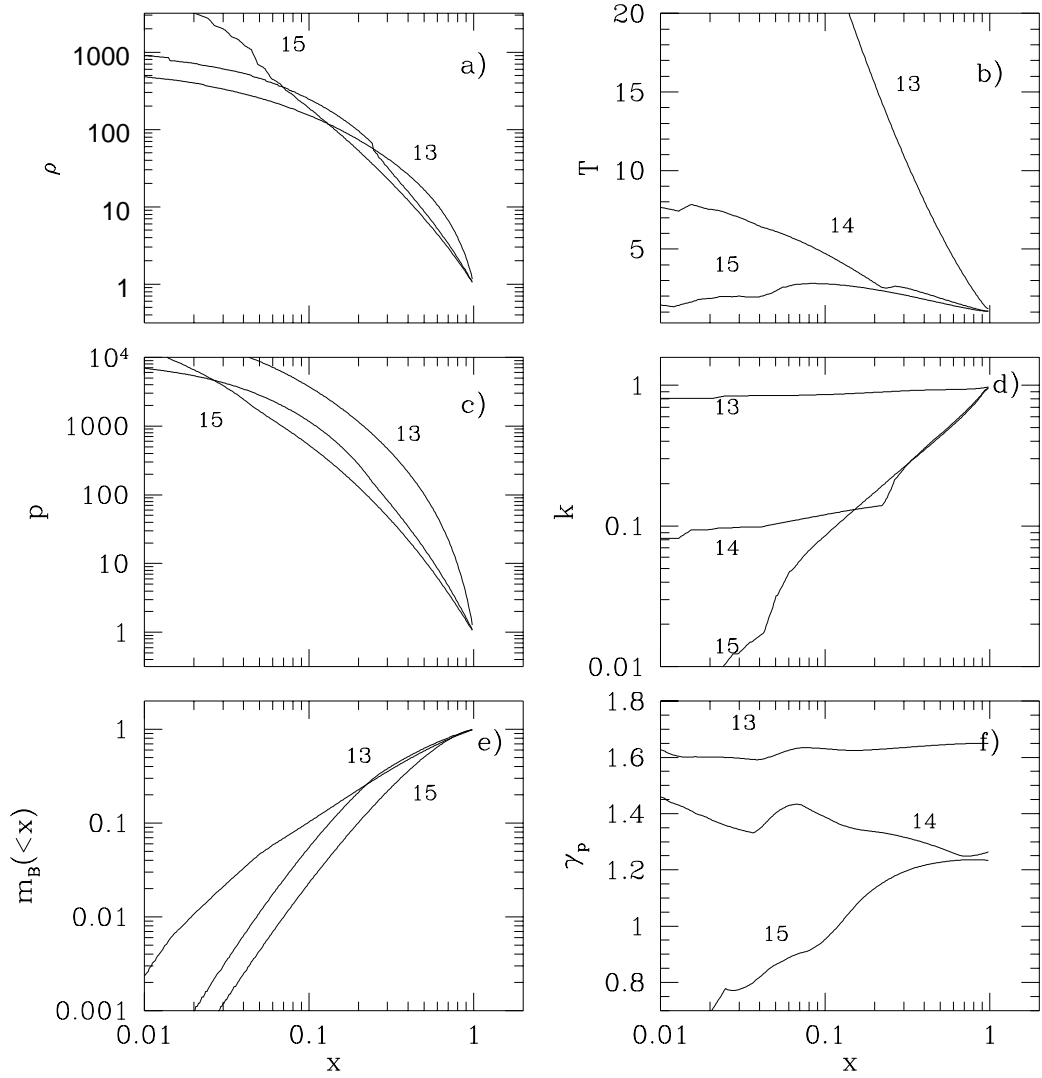,height=15truecm,width=15truecm}}
\caption{The same as in Figure \ref{fig5} for $\Lambda$CDM, but with the
inclusion of cooling; the external, initial adiabat is $K_* =0.3
\times 10^{34}$ erg cm$^2$ g$^{-5/3}$.  Note in panel d) that the
entropy plateau in the center has been partially erased by cooling.
The polytropic indexes $\gamma_p$ are averaged over $\Delta
{\rm log}(x) = 0.3$.
\label{fig7}}
\end{figure}

Figure \ref{fig7} shows the case with $K_{34}=0.3$ and with the
inclusion of cooling.  This case can be considered a realistic,
complete scenario.  As we shall see later, this value of the
background entropy gives a good fit to the $L$--$T$ relation.  The
inclusion of cooling introduced some change with respect to Figure
\ref{fig5}, especially in the very inner regions, where the entropy
evolved towards lower values.  However the entropy excess in the
center is still present (panel d). Cores with constant density appear
more peaked, but small groups still show much flatter density profiles
with respect to large clusters.  The temperature profiles are lower,
and the polytropic index $\gamma_p$ is rapidly decreasing in the
center.

\begin{figure}
\centerline{\psfig{figure=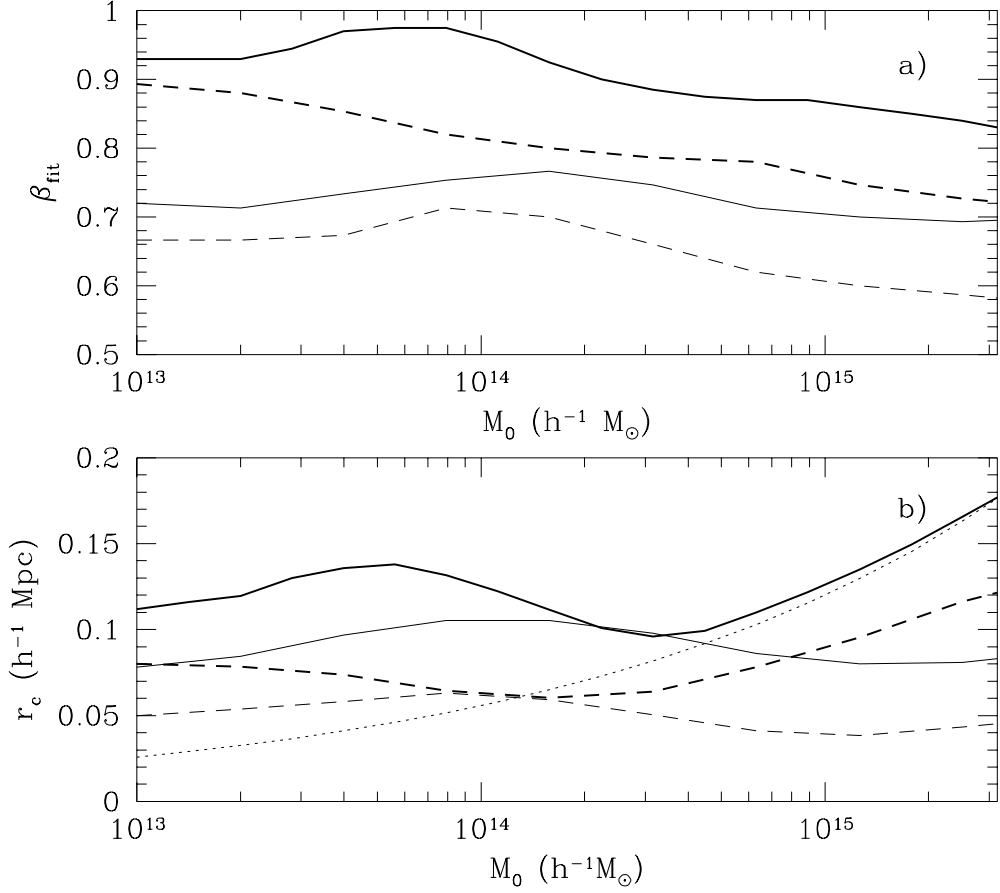,height=15truecm,width=15truecm}}
\caption{--$\beta_{fit}$ parameter and the core radius $r_c$ as a
function of the mass scale derived by fitting the predicted profiles
with a beta model.  The initial entropy is $K_* =0.3 \times 10^{34}$
erg cm$^2$ g$^{-5/3}$, constant with epoch, and the cooling is
included.  The thick lines refer to $z=0$ and the thin lines to $z=1$.
The $\Lambda$CDM universe is shown with solid lines, while tCDM
with dashed lines.  A self--similar scaling radius ($r\propto
M^{1/3}$) is shown with a dotted line for comparison.
\label{fig8}}
\end{figure}

For a more comprehensive view, the differences in the density profiles
can be expressed in terms of fitting parameters $\beta$ and $r_c$
after adopting a beta model (Cavaliere \& Fusco Femiano 1976).  The
results are shown in Figure \ref{fig8}. The $\beta_{fit}$ parameter is
about $\simeq 0.8$ in $\Lambda$CDM at $z=0$, and about $0.6$ at $z=1$.
The density profiles are slightly steeper in the outer regions at
smaller masses.  However the most prominent feature is the core
radius, whose scaling departs from the self--similar behaviour
$R\propto M^{1/3}$ (dotted line) below $1$ keV.  No significant
differences are predicted in the tCDM universe.  The flattening of the
$R-M$ or the $R-T$ relation has been clearly detected in the data, and
related to heating processes, by Mohr \& Evrard (1997); note, however,
that they plotted an isophotal radius, which is a much better defined
quantity from the observational point of view.  Smaller cores are
found at higher $z$, since all the linear dimensions are reduced
approximately by a factor $(1+z)$.

We note that our results differ from those found by Fujita \& Takahara
(2000).  In fact, their assumption of isothermality allows to relate
the $\beta$ parameter directly to the temperature of the external gas.
This is no longer valid in our model, where the entropy of the
external gas affects the dimension of the adiabatic core $r_c$,
breaking the self--similar scaling, while yielding a $\beta$ parameter
independent of the mass.

Finally, we note that our values of $\beta$ are somewhat larger than
that observed in clusters (see Mohr, Mathiesen \& Evrard 1999).  This
may be due to our fitting procedure, that extends up to the shock
radius.  In fact, our profiles are steeper than a $\beta$--model at
large radii, and the best fits usually give larger $\beta$ for larger
cores in order to reproduce the rapid steepening of the profiles out
of the core.  The outer regions are generally too weak to be detected
in ROSAT data, since their surface brightness is below
$10^{-15}$ erg s$^{-1}$ cm$^{-2}$ arcmin$^{-2}$ (see TSN00).  On the
other hand, such a regions are expected to be efficiently detected in
the future Chandra and XMM data.

\subsection{The effect of Cosmology and Dark Matter}

\begin{figure}
\centerline{\psfig{figure=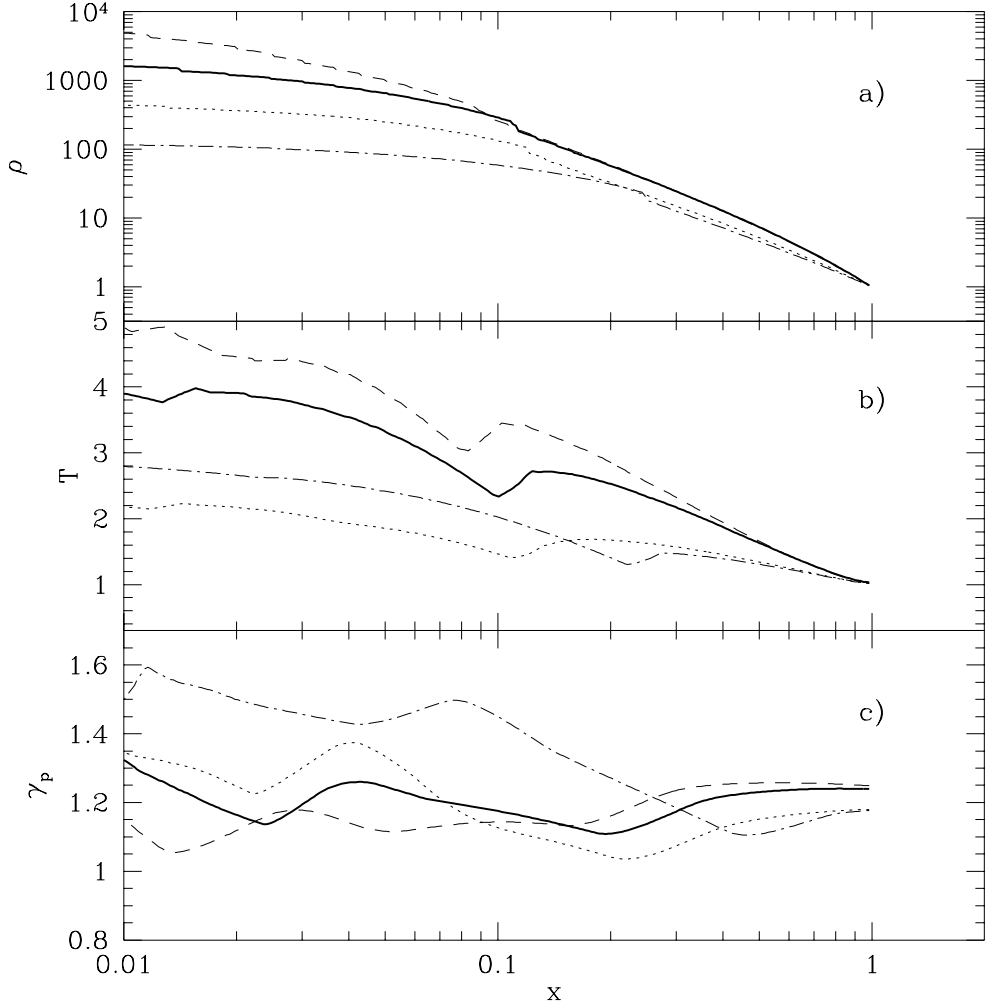,height=15truecm,width=15truecm}}
\caption{The heavy solid lines show the normalized density (panel
a) and the normalized temperature (panel b) profiles for a cluster of
$M=0.6 h^{-1} 10^{15} M_\odot$ ($k_BT\simeq 5$ keV) in $\Lambda$CDM
starting from the external adiabat $K_* =0.4 \times 10^{34}$ erg
cm$^2$ g$^{-5/3}$ (cooling is included).  For comparison, we show the
effect of changes in the dark matter profile from NFW to Moore et
al. 1998 (dashed lines), of the epoch from $z=0$ to $z=1$ (dot--dashed
lines) and of cosmology from $\Lambda$CDM to tCDM (dotted line).  In
panel c) the corresponding polytropic indexes $\gamma_p$ are shown
(averaged over $\Delta {\rm log}(x) = 0.3$).
\label{fig9}}
\end{figure}

From the considerations above, it is clear that the level of the
initial adiabat strongly affects the final properties of the ICM, and
that, in principle, it is not necessary to invoke substantial heating
after the collapse, provided that $K_{34}\simeq 0.3$.  The profiles
are affected also by changing the cosmological background, the epoch
of observations, or the dark matter profile.  To show these
variations, not directly related to the entropy, in Figure \ref{fig9}
we plot the density and temperature profiles, along with the
polytropic index, for a typical massive cluster ($0.6 \times 10^{15} ~
h^{-1}~M_\odot$, corresponding to a virial temperature of $k_BT\simeq
5$ keV) changing in turn cosmology, epoch and dark matter profile and
comparing them to the case with $\Lambda$CDM at $z=0$, NFW profile,
$K_{34} =0.4$.  The cooling is included for all the cases.

A steeper dark matter profile (Moore et al. 1998) gives higher gas
densities in the center (dashed line).  The temperature gradient is
correspondingly larger.  In any case $\gamma_p$ is always bounded
between $0.9$ and $ 1.2$ outside the adiabatic core.  In principle,
observations can discriminate between different dark matter profiles,
and the observed temperature profiles (see Markevitch 1998) would
favour profiles steeper than NFW, but we recall that this minor effect
can be overwhelmed by changes in the entropy or by the presence of
substructure.

At higher redshifts (here we focus on a typical value of $z=1$ which
is the nominal goal of the future X--ray surveys) the adiabatic
accretion is relatively more extended in time during the lifetime of
the object, and, for the same value of $K_*$, the imprint of the
background entropy is more evident.  This is because virialized
objects form at a total density contrast which is almost constant with
respect to the critical density, and the baryons will consequently
reach larger densities before being accreted.  These larger densities
translate into pre--shock temperatures larger approximately by a
factor $(1+z)^2$, and thus in a larger sound speed $c_s\propto (1+z)$.
On average, the shock condition is harder to satisfy since the
infalling velocities scale only as $v_i\propto \sqrt{1+z}$, and,
consequently, a larger number of baryons are accreted adiabatically.
The resulting density and temperature profiles are flatter (dot--short
dashed line).  This effect adds to the flattening of the total dark
matter profile at high redshift, as envisaged by NFW.  As we will see,
this mechanism is responsible for keeping the $L$--$T$ relation
approximately constant with redshift.

The case for a tCDM cosmology at $z=0$ (dotted line) shows flatter
profiles.  This is easily understood if we recall that the external
density is proportional to the mass accretion rate, and that the mass
accretion rates are higher at $z=0$ in tCDM with respect to
$\Lambda$CDM (similar to the rates at $z=1$ in $\Lambda$CDM for
objects of the same mass).  In general, the cosmology does not have a
large effect on the evolution of the $L$--$T$.

\subsection{The Shock Radius and the Baryonic Fraction}

The boundaries of the emitting gas are given by the shock radius of
the last accreted shell, where there is a discontinuity between the
inner hot gas and the outer cooler gas.  The outer unshocked gas gives
also a contribution to the emission, and can be detected in the
outskirts of rich clusters giving important information on the entropy
level of the external baryons (TSN00).  It always gives a small
contribution if compared to the total emission from the cluster, and
here it is neglected.  For very small mass objects, the last accretion
radius is quite distant from the virial radius, in a region of very
low density and very low infall velocity.  The shocks are typically
very weak, and the gravitational entropy production is negligible.  In
such low mass objects the X--ray emission is expected to fade outwards
without discontinuity.

\begin{figure}
\centerline{\psfig{figure=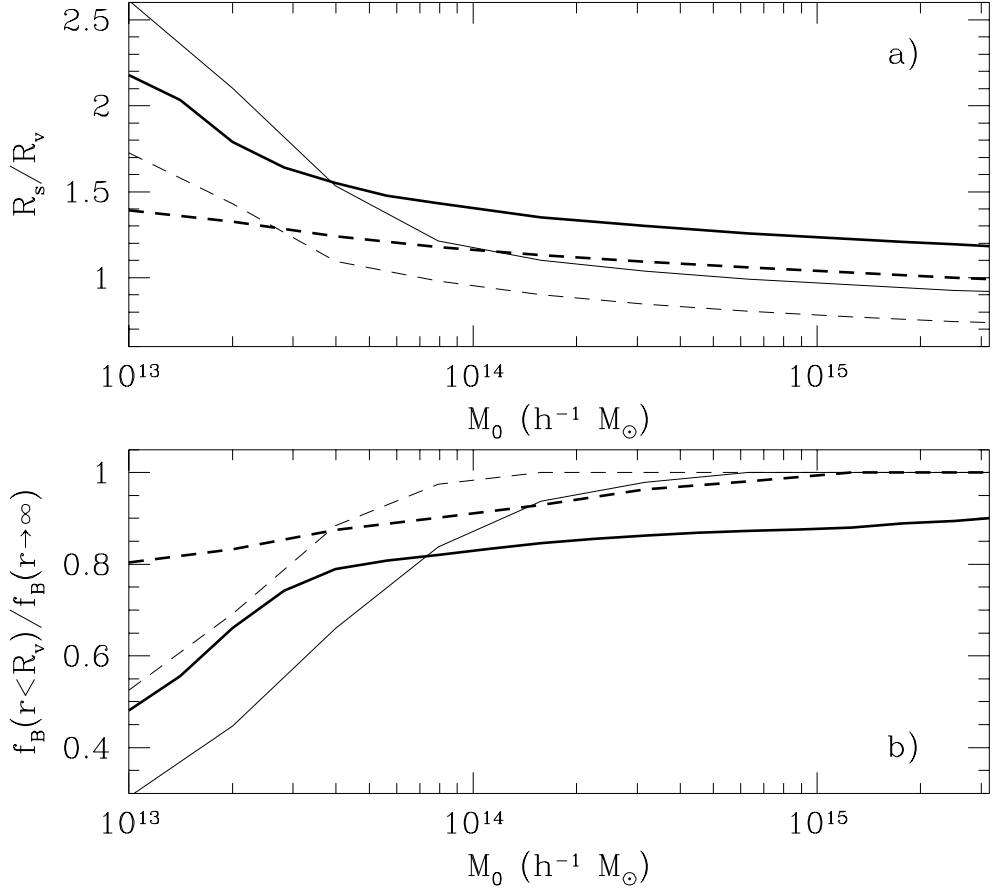,height=15truecm,width=15truecm}}
\caption{a) Ratio of the shock radius of the last accreted shell
to the virial radius as a function of the mass scale; b) the baryonic
fraction (with respect to the universal baryonic fraction) within the
virial radius as a function of mass.  The background entropy is $K_* =0.3
\times 10^{34}$ erg cm$^2$ g$^{-5/3}$ constant with epoch, and cooling
is included.  The thick lines refer to $z=0$ and the thin to $z=1$.
The $\Lambda$CDM universe is shown with solid lines, while tCDM
with dashed lines.  
\label{fig10}}
\end{figure}

The position of the last accreting shell is calculated simply using
mass conservation.  In fact, following Equation (\ref{mdot}), the
total mass of diffuse baryons involved in the cluster collapse, is
equal to the mass included in the initial comoving region, $M_B = f_B
M$, after subtraction of the baryons in stars and the cooled baryons
in the center which depend on epoch and mass.  Due to the different
distribution of the baryons with respect to the dark matter, the ratio
of the shock to the virial radius is a function of epoch and of the
total mass accreted, as shown in Figure \ref{fig3}.  In Figure
\ref{fig10}(a) we show the position of the final shock radius
with respect to the virial one at redshift $z_0=0$.  At small masses,
where the gas distribution is flatter and more extended, the shock
radius can be approximately $\simeq 2$ times larger than the virial
radius.  In other words, the external gas does not fall into the
potential well, but is accumulated at large radii.  At very high
masses the accretion rates are larger, and the pressure term can be
important, giving, for high density universes, a shock radius slightly
smaller than the virial radius. The same happens at higher redshift
when the accretion rates are correspondingly larger. In any case, for
large mass systems the shock radius is expected to remain close to the
virial radius of the cluster, as was predicted in numerical
simulations (see, e.g., Takizawa and Mineshige 1998).  Slightly larger
shock radii are predicted for higher values of the background entropy.

Here, the ratio of the mass in baryons within the shock radius to the
total mass within the virial radius, is, by definition, always equal
to the universal average baryonic fraction.  However, since the two
radii are generally different, the observed baryonic fraction within
the virial radius will be a growing fraction of the mass scales.  In
Figure \ref{fig10}(b) the baryonic fraction within the virial
radius $R_v$ is shown as a function of the total virialized mass. The
largest variations are between masses $10^{13}$ and $10^{14} h^{-1}
M_\odot$, roughly corresponding to temperatures below $1$ keV at
$z=0$.  In any case, any entropy excess, irrespective of the origin
(external or internal) always tends to puff up the baryons with
respect to the dark matter (as observed in numerical simulations, see
Pearce et al. 1994, Tittley \& Couchman 2000).  This reinforces the
case for a low density universe derived from the observed high ratio
of baryonic to total mass.

\subsection{The Energy Budget}

An important quantity is the amount of non--gravitational energy per
particle corresponding to $K_*$.  The temperature corresponding to a
given adiabat is $k_BT\simeq 3.2\times 10^{-2}\, K_{34} (1+z_h)^2
\delta^{2/3}$ keV, where $\delta$ is the overdensity with respect to
the ambient density at $z_h$, and $z_h$ is the epoch of the heating.
The assumption of an initial and homogeneous $K_*$, implies that the
entropy of each shell must be in place at turn--around.  At this epoch
the density of the shell is assumed to be the background value .
Thus, the minimum energy released in the gas can be computed as
\begin{equation}
 k_BT_{min} = 3.2 \times 10^{-2} K_{34}{ 1 \over M_0} \int^{M_0}_0
[1+z_{ta}(M)]^2 dM\, .
\end{equation}
In the case of a $\Lambda$CDM universe we have $k_BT_{min} \simeq 0.1
(K_{34}/0.4)$ keV with a small dependence on the final mass $M_0$.  

As we have already discussed, starting from a high adiabat is not the only
way to prevent massive cooling, since non--gravitational heating in
the center of the clusters could help in re-establishing the entropy
floor.  However, the energy needed to re--establish the entropy floor
after accretion is much higher than the energy needed to put the
baryons initially on the {\sl right adiabat}.  If the baryons are
heated preferentially at higher density the excess energy is higher by
a factor $\delta^{2/3}$.  However, this is not the only reason for a
larger energetic budget.  In fact, another advantage in heating the
gas at lower densities, is that radiative cooling is not able to
re--emit the energy on very short time--scales.

To make a simple example without the cooling, if the baryons are
heated at $z\simeq 0$ when they are at an average density contrast
equal to ~$200$, typical of virialization, we would obtain $k_BT\simeq
0.3 $ keV.  However, this value underestimates the real energy budget,
since the density in the center, where the entropy excess is expected,
is much higher than the average contrast, and $z\simeq 0$ is in any
case too late to inject the extra energy.  A more realistic
calculation for the center of rich clusters can require more than $2$
keV per particle (see \S 7 and TSN00) to establish a density core and
eventually halt the cooling in the center.  This arguments show
clearly how the same entropy level, which determines all the X--ray
emission properties, can be due to very different heating balances.
In this respect, the distribution of metallicity in the ICM may be
useful in calculating the actual amount of excess energy dumped into
the baryons.

\subsection{The Luminosity--Temperature--Mass relations and the 
Entropy--Temperature plot}

We can derive the average relation between the bolometric
luminosity, the emission weighted temperature and the total virialized
mass.  The bolometric luminosity over the whole emitting volume
defined by $R_s$ is:
\begin{equation}
L_x = \int^{R_s}_0 \epsilon (r) dV \, \, {\rm erg \, s^{-1}}\, ,
\label{lum}
\end{equation}
where $\epsilon(r)$ is the emissivity per unit volume, including
free-free and line emission, expressed by:
\begin{equation}
\epsilon = n_e n_i \Lambda_N\, \, {\rm erg
\, s^{-1} \, cm^{-3}}
\end{equation}
where $n_e$ and $n_i$ are the electron and ion density respectively,
and $\Lambda_N$ is the normalized cooling function depending on
temperature and metallicity (from Sutherland \& Dopita 1993). We adopt
a value of $Z = 0.3\, Z_\odot$, as observed on the scale of clusters
($k_BT>2$ keV).  Such a value is currently observed on the scale of
groups with large uncertainties, due to difficult line diagnostic
and poor temperature resolution (Renzini 1997; Buote 2000).  However,
since the cooling function includes emission over a range of energies
wider than the usual X--ray bands, we cut the emission at energies
lower than $0.1$ keV.

The emission weighted temperature defined over the entire emitting
volume is:
\begin{equation}
k_BT_{ew}\equiv {{\int^{R_s}_0 k_BT(r)\epsilon (r)
dV}\over{\int^{R_s}_0}\, \, \epsilon (r) dV}\,{\rm keV}.
\label{temp}
\end{equation}

\begin{figure}
\centerline{\psfig{figure=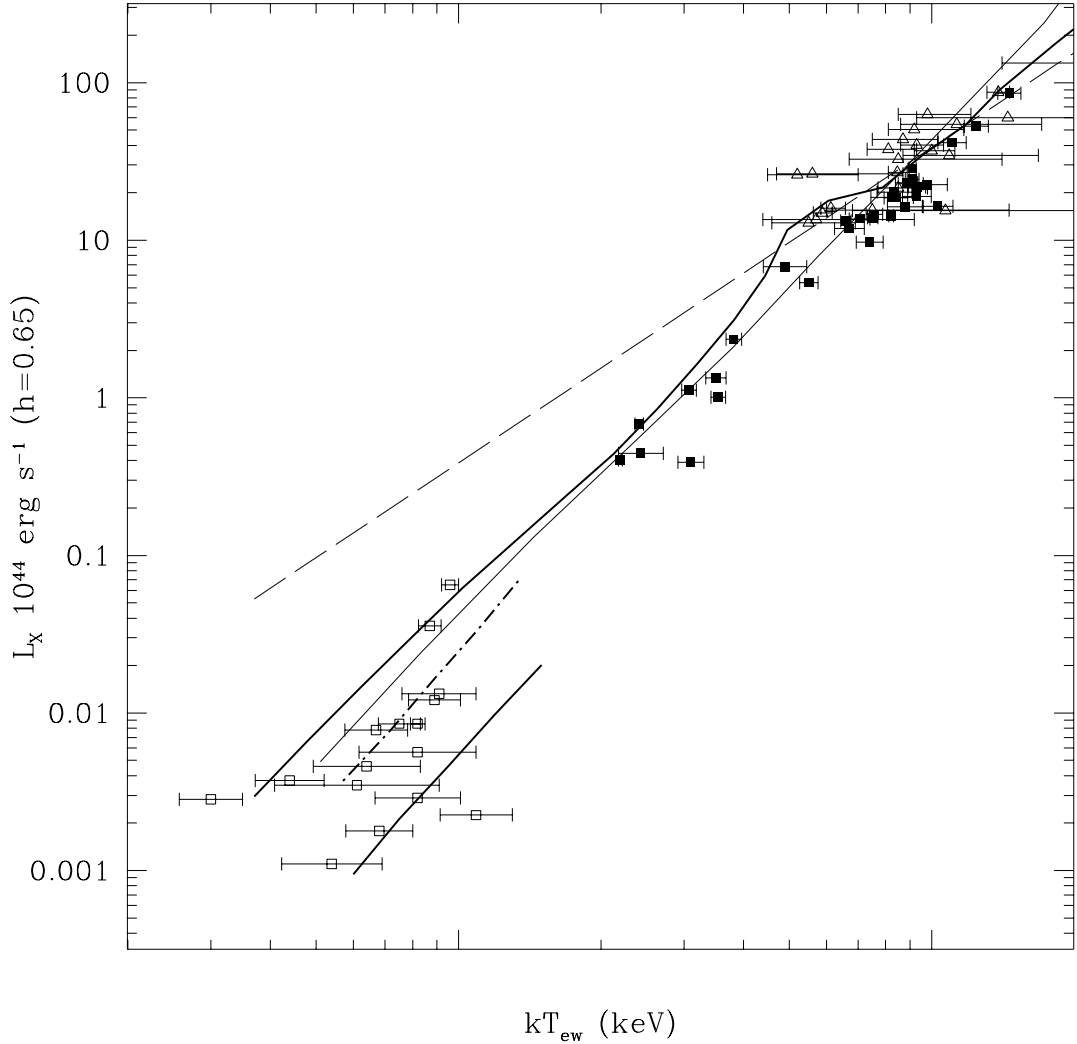,height=15truecm,width=15truecm}}
\caption{Relation between bolometric luminosity and emission
weighted temperature in $\Lambda$CDM. The entropy background is
$K_*=0.3 \times 10^{34}$ erg cm$^2$ g$^{-5/3}$ constant with epoch,
and the cooling is included.  Data are from Arnaud \& Evrard (1999,
filled squares), Allen \& Fabian (1998, triangles) and Ponman et
al. (1996, empty squares).  The dashed lines refer to the
self--similar case, while thick lines to $z_0=0$ and thin lines to
$z_0=1$.  The lower thick solid line for $k_BT\leq 1.5$ keV shows
the $L$--$T$ relation at $z=0$ defined within the projected radius of
$100 \, h^{-1} $ kpc as in Ponman et al. (1996).  The lower thick
dot--dashed line shows the same for $K_*=0.2 \times 10^{34}$ erg
cm$^2$ g$^{-5/3}$.
\label{fig11}}
\end{figure}

\begin{figure}
\centerline{\psfig{figure=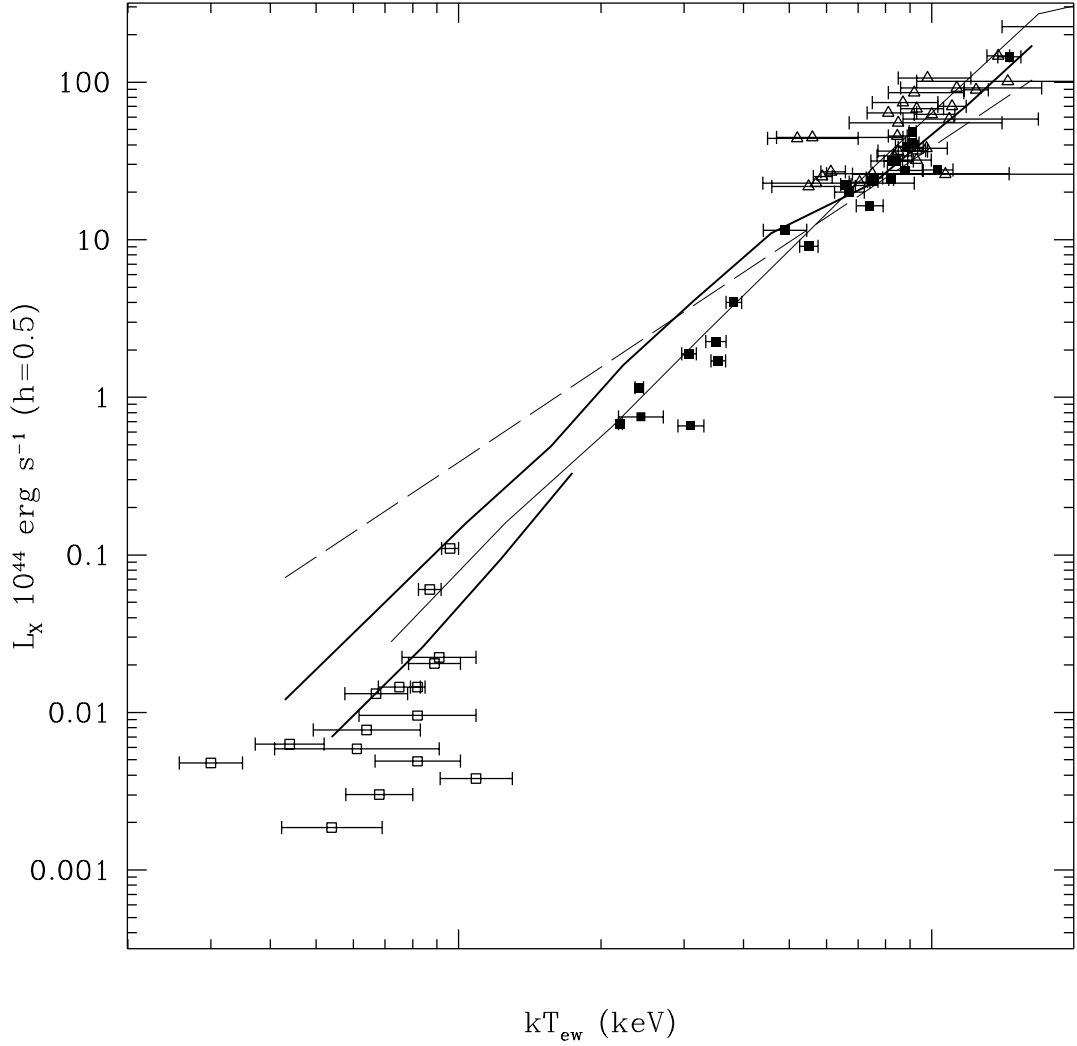,height=15truecm,width=15truecm}}
\caption{Relation between bolometric luminosity and emission
weighted temperature in tCDM, notations as in Figure \ref{fig11}.
Note that we are forced to use a baryonic density $\Omega_B = 0.04
h^{-2}$, which is at least twice the value from standard
nucleosynthesis constraints.  The lower thick line for $k_BT\leq 1.5$
keV shows the $L$--$T$ relation at $z=0$ defined within the projected
radius of $100 \, h^{-1} $ kpc as in Ponman et al. (1996).  Data as in
Figure \ref{fig11}.
\label{fig12}}
\end{figure}

The results are shown in the Figures \ref{fig11} and \ref{fig12} for
$\Lambda$CDM and tCDM respectively for $K_{34} = 0.3$, with the
inclusion of cooling.  The self--similar case is shown for comparison
(dashed line).  Data are taken from Arnaud \& Evrard (1999) and Allen
\& Fabian (1998) for the clusters, and from Ponman et al. (1996) for
the groups.  An important issue here is that the total luminosity
emitted by all the accreted gas (light curves in Figures \ref{fig11}
and \ref{fig12}), overestimates the luminosities found by Ponman et
al. (1996) at temperatures below $1$ keV.  This is because the
luminosities of the observed groups are defined within the fixed
projected radius of $100 h^{-1} $ kpc.  Therefore we also calculated
the luminosity and the emission weighted temperature performing the
integrals of Equations (\ref{lum}) and (\ref{temp}) over the
cylindrical volume defined by the projected radius of $100h^{-1}$ kpc.
We show both the total luminsity, including all the gas even at
$R_s>>R_v$, and the luminosity within $100h^{-1}$ kpc.  The lower
values with respect to the global $L$--$T$ relation is due to a factor
of $\simeq 1/3$ in luminosity due to the exclusion of the low
surface--brightness gas at radii larger than $100h^{-1}$ kpc, and by
the factor of $\leq 2$ gained in the emission weighted temperature
since only the inner regions, with strong temperature gradients, are
included.  Thus, in the simple scenario of an external $K_*$, the
groups are expected to be surrounded by a large halo of
surface--brightness $\simeq 10^{-16}$ erg s$^{-1}$ cm$^{-2}$
arcmin$^{-2}$.  Its detection would constitute an important test for
the external entropy scenario (TSN00).

In Figure \ref{fig11} we also show the prediction for the luminosity
within $100h^{-1}$ kpc in the cases $K_{34}=0.2$, which turns out to
give better fits for the groups.  Thus, even if clusters with $k_BT>2$
keV seems to require $K_{34}\simeq 0.3-0.4$, a lower value $K_* \simeq
0.2$ gives a better fit to the low end of the $L$-$T$ relation.  As we
will see, this is confirmed by the entropy--temperature relation (see
Figure \ref{fig17}).

It is clear how the presence of the background entropy bends the
$L$--$T$ relation from the self--similar slope to an average $L\propto
T^3$.  However, with this simple model it is difficult to reproduce
the steepening below $1$ keV.  This is partially due to inclusion of
line emission, that prevents the $L$--$T$ relation from reaching the
adiabatic slope $L\propto T^5$.  In fact, for a metallicity
$Z>0.1Z_\odot $ the slope of the emission curve between $0.3-1$ keV is
virtually zero, or even negative.  In this case the asymptotic slope
will be flatter than $T^{4}$.

\begin{figure}
\centerline{\psfig{figure=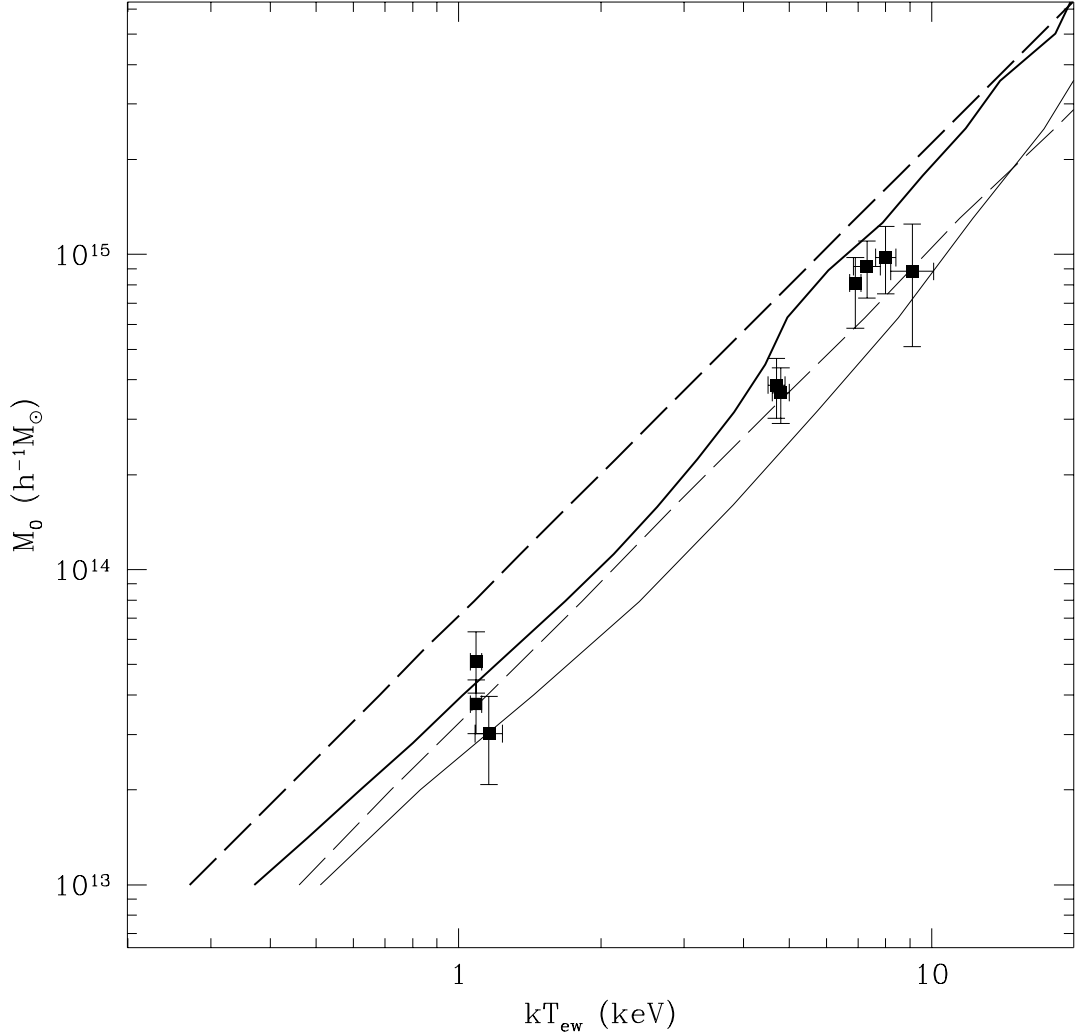,height=15truecm,width=15truecm}}
\caption{Relation between virial mass and emission
weighted temperature in $\Lambda$CDM; data from Nevalainen, Markevitch
\& Forman (2000). The entropy background is $K_*=0.3 \times 10^{34}$
erg cm$^2$ g$^{-5/3}$ constant with epoch, and the cooling is
included.  The dashed lines refer to the self--similar case, while
thick lines to $z_0=0$ and thin lines to $z_0=1$.  Note that the mass
has been rescaled from $M_{500}$ to the virial value using the
corresponding NFW profile.
\label{fig13}}
\end{figure}

\begin{figure}
\centerline{\psfig{figure=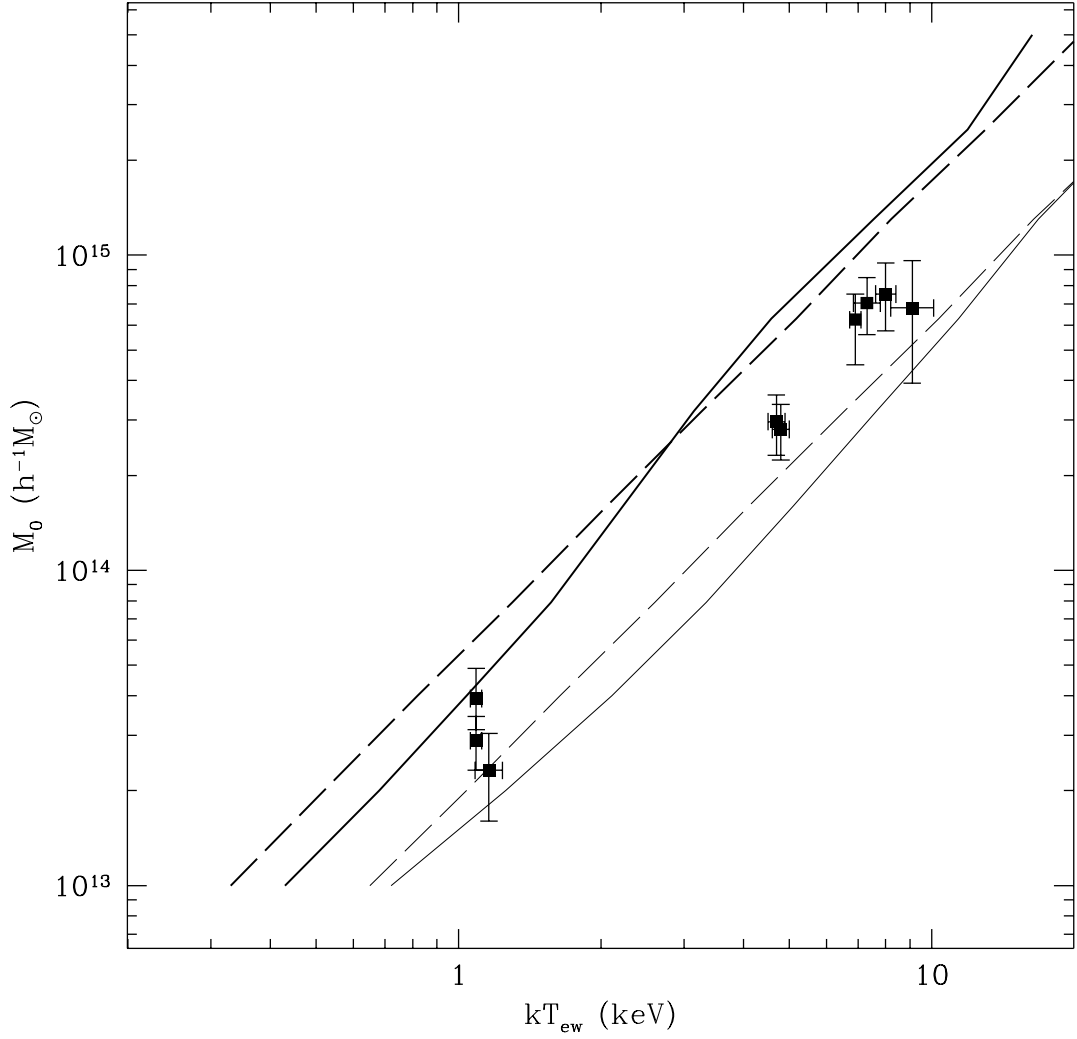,height=15truecm,width=15truecm}}
\caption{Relation between emission weighted temperature and virial
mass in tCDM. The entropy background is $K_*=0.3 \times 10^{34}$ erg
cm$^2$ g$^{-5/3}$ constant with epoch, and the cooling is included.
The dashed lines refer to the self--similar case, while thick lines to
$z_0=0$ and thin lines to $z_0=1$.  Data as in Figure \ref{fig13}.
\label{fig14}}
\end{figure}

The $M$--$T$ relation at small masses is lower with respect to the
relation between mass and virial temperature (dashed lines, see
Equation (2.2) in Eke et al. 1998) which is reproduced by our
self--similar case.  The predicted $M$--$T$ relation in $\Lambda$CDM
with $K_{34} = 0.4\pm 0.2$ is consistent with the recent finding of
Nevalainen, Markevitch \& Forman (2000).  Note that the values plotted
in Figures \ref{fig13} and \ref{fig14} are re-scaled to the virial mass
from the mass quoted in the paper, using the corresponding NFW
profile.  The steepening of the temperature profiles in the adiabatic
cores gives higher emission weighted temperatures, about $\simeq 25$\%
larger than the corresponding virial temperatures for $k_BT<2$ keV.
This translates into an uncertainty of less than a factor of $2$ in
the total mass (using the self--similar relation).  The evolution is
similar to that of the self--similar case, and the difference in slope
is preserved.  In the tCDM case, the $M$--$T$ relation is higher and
gives a poor fit to the data of Nevalainen, Markevitch \& Forman
(2000).

\begin{figure}
\centerline{\psfig{figure=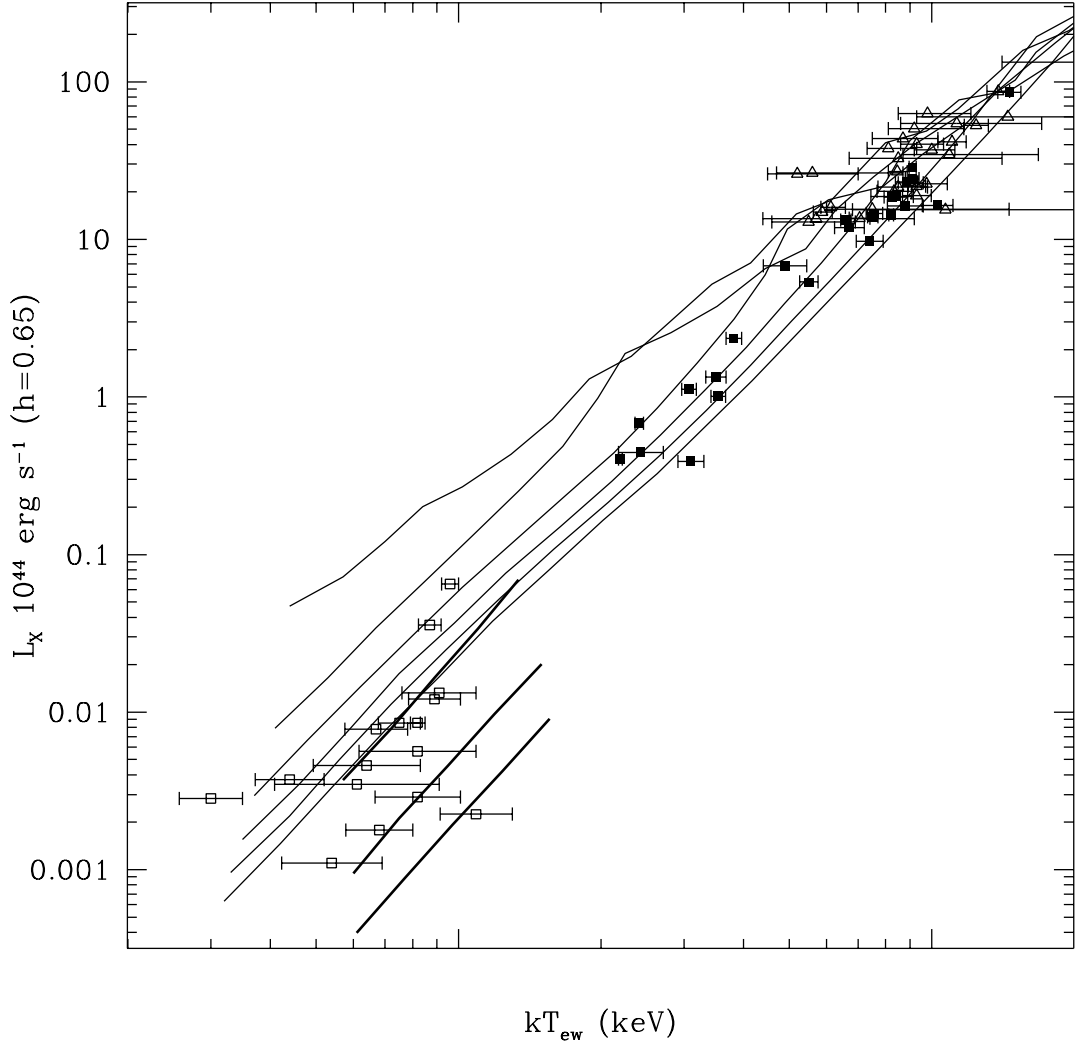,height=15truecm,width=15truecm}}
\caption{The thin lines show the $L$--$T$ relation at
$z=0$ in $\Lambda$CDM for different values of the initial adiabat
(background entropy): from top to bottom $K_* = 0.1 - 0.2 - 0.3 - 0.4 -
0.5 - 0.6 \times 10^{34}$ erg cm$^2$ g$^{-5/3}$.  The thick segments
shows the prediction limited to the inner $100 h^{-1}$ kpc for $K_* =
0.2 - 0.3 - 0.4 \times 10^{34}$ erg cm$^2$ g$^{-5/3}$.  Data as in Figure
\ref{fig11}.
\label{fig15}}
\end{figure}

\begin{figure}
\centerline{\psfig{figure=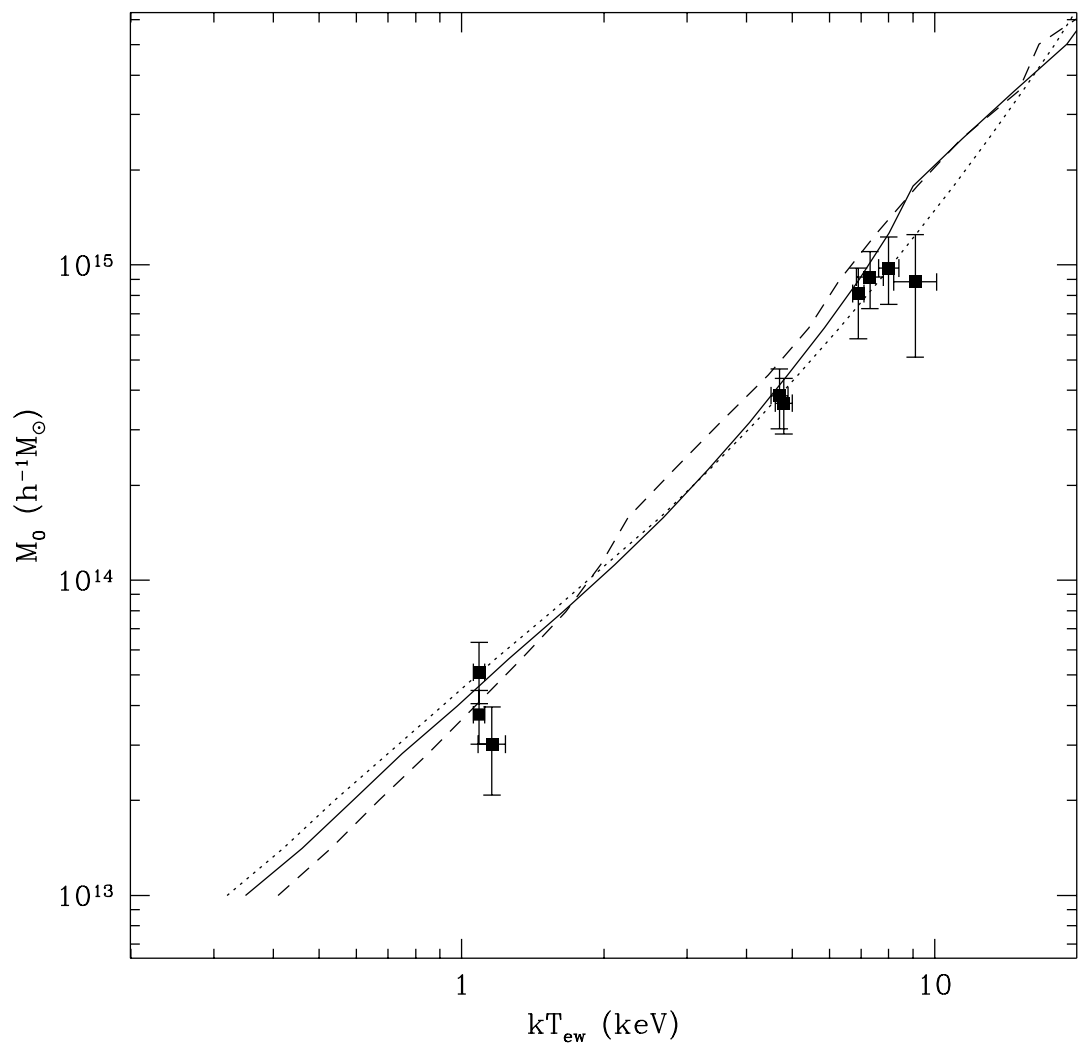,height=15truecm,width=15truecm}}
\caption{The $M$--$T$ relation in $\Lambda$CDM for
different values of the initial adiabat: $K_* = 0.2 - 0.4 - 0.6 \times
10^{34}$ erg cm$^2$ g$^{-5/3}$ (respectively dashed, solid and dotted
lines).  Data as in Figure \ref{fig12}.
\label{fig16}}
\end{figure}

\begin{figure}
\centerline{\psfig{figure=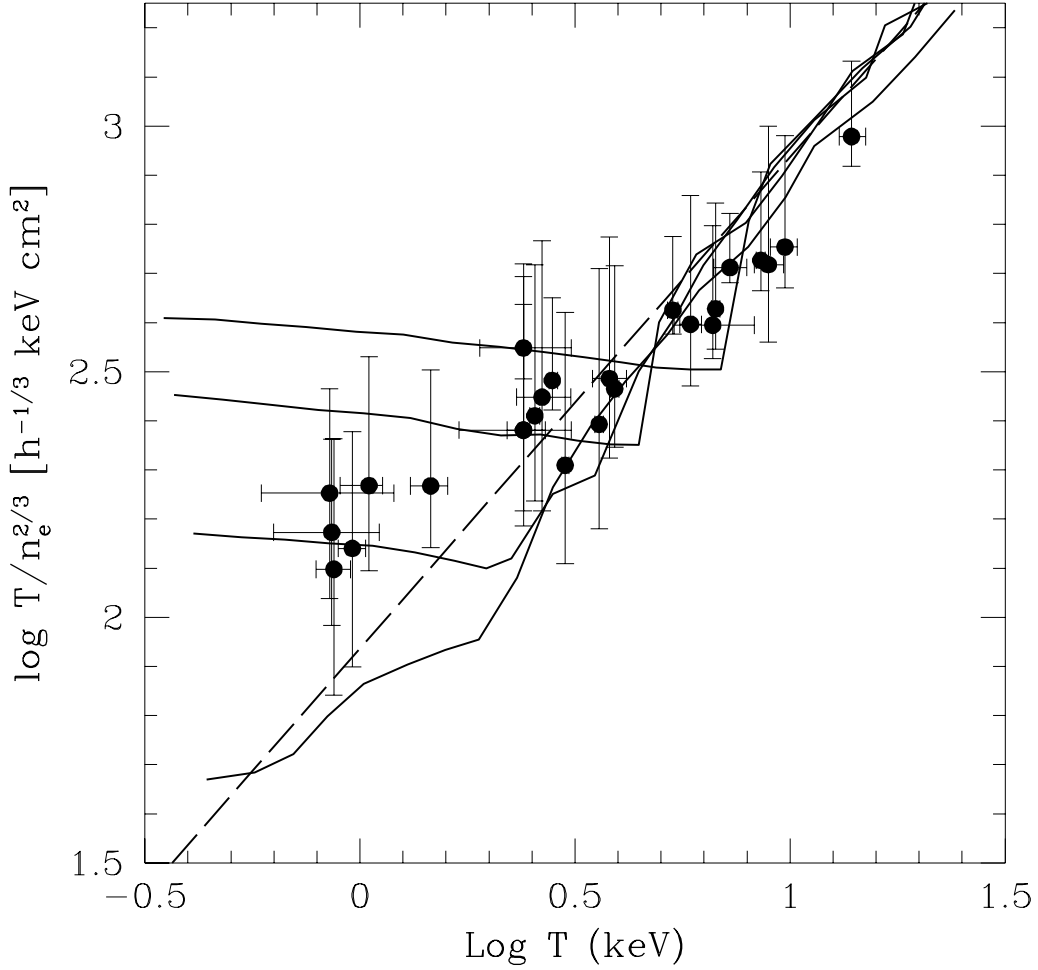,height=15truecm,width=15truecm}}
\caption{Relation between central entropy defined as in PCN as
$T/n_e^{2/3}$ at $r=0.1R_v$, as a function of the local temperature
$T(r)$ for $\Lambda$CDM at redshift $z=0$ for different value of
the background entropy $K_* = 04. - 0.3 - 0.2 - 0.1 \times 10^{34}$ erg
cm$^2$ g$^{-5/3}$ (from top to bottom).  Data are from PCN. The dashed
line is the self similar case from N--body simulations, after PCN.
\label{fig17}}
\end{figure}

The slope of the $L$--$T$ relation is affected by different values of
$K_*$ as shown in Figure \ref{fig15} at $z=0$ in a $\Lambda$CDM
universe.  Lower values gradually approach the self--similar relation
$L\propto T^2$.  However, the self--similar scaling is never reached in
the limit $K_*\rightarrow 0$, due to the cooling catastrophe.  We
recall that our ability to include the cooling processes in the cases
presented here, is due to the non--negligible initial entropy level.
If $K_{34} <<0.05$, the cooling processes are too strong and our
computation scheme becomes inadequate.  The $M$--$T$ relation is less
affected by changes in $K_*$ (see Figure \ref{fig16}).

All the above physics influences the relation between the central
entropy (measured at a radius $r=0.1R_{v}$) and the temperature, as
shown in PCN.  The emergence of the entropy floor at small scales (low
temperatures) is directly seen as a departure from the self--similar
expectations, shown as a dashed line in Figure
\ref{fig17}\footnote{The entropy is computed using the predicted local
value of the temperature at $r=0.1R_{v}$; very similar values are
obtained using the emission weighted temperature as effectively used
in PCN}.  Note that in this case the adiabat is defined differently,
using the electron density instead of the mass density: $K_P\equiv
k_BT/n_e^{2/3}$ keV cm$^2$.  The relation between the two definitions is
$K_P = 0.95\times 10^3 K_{34}$.  In this respect, the value observed
should be considered indicative of the average entropy in the center
of the halos.  The entropy floor is clearly matched at $k_BT< 2$ for
$0.1< K_{34}< 0.4$.  In particular, $K_{34}=0.2-0.3$ reproduce both
the $L$--$T$ and $K$--$T$ relations over the whole temperature
range.

\section{THE ENTROPY HISTORY OF THE UNIVERSE AND THE X--RAY EVOLUTION
OF CLUSTERS}

From the above results, it is clear that a significant background
entropy, $S_*$, present in the IGM before the formation of large dark
matter halos affects the X--ray properties of groups and clusters and
can explain many scaling properties.  However, the assumption of a
uniform floor of entropy for all the baryons could be too simplistic.
As we showed, the data seems to require a growing value of $K_*$ at
larger mass scales: $K_{34}\simeq 0.2$ for $k_BT <2$ keV, and
$K_{34}\simeq 0.4$ for $k_BT >2$ keV.  In terms of physical
mechanisms, it is reasonable to expect that $S_*$ is correlated with
higher density regions where star formation or nuclear activity
preferentially occurs.  For example, if the excess entropy is linked
to star formation processes, an entropy excess should be observed in
the diffuse baryons expelled by galaxies at high redshift.  The
distribution of entropy should follow the light distribution, and
should show a dependence on cosmic time that parallels the birth of
the first stars and QSOs. This topic can be addressed not only with
X--ray observations, but also with the UV and optical investigation of
the low density baryons detected, e.g., as Ly$\alpha$ clouds.  Here we
will discuss in greater detail the scenario with a uniform external
entropy, but relaxing the assumption of a constant $K_*$.

We already know that the IGM which is observed in high--$z$ Ly$\alpha$
clouds generally shows an entropy level lower than that observed in the
centers of groups.  An approximate relation derived from the
observations is $K_{Ly\alpha} = (1.2\pm 0.5) \, 10^{-2}
(1+z)^{-1}\times 10^{34}$ erg g$^{-5/3}$ cm$^2$ (extrapolated from
Figure 10(b) in Ricotti et al. 2000, see also Schaye et al. 1999).
Thus, the ratio of the value $K_{gr}$ observed in the center of the
groups to that observed in Ly$\alpha$ is about
$K_{gr}/K_{Ly\alpha}\geq 10 (1+z)$. This may indicate that the ICM
baryons undergo substantial heating with respect to the baryons
observed in Ly$\alpha$ or, possibly, that the baryons seen in
Ly$\alpha$ clouds are not the {\sl same} baryons that will be later
accreted in clusters.  Furthermore, the chemical properties of the IGM
seen in the Ly$\alpha$ forest are clearly different from those of the
ICM in clusters, showing that the ICM was affected by star formation
processes and chemical enrichment to a larger extent with respect to
the Ly$\alpha$ clouds, with a commensurate amount of entropy
production.  In this respect, it will be interesting to observe the
tenuous gas being accreted in the outskirts of nearby, large clusters,
but not yet shocked, or at large radii in small groups, and compare it
with the gas observed in different enviroments at different cosmic
epochs.  Such observations would complement the investigation of the
entropy excess as observed in nearby and distant clusters.

As expected, the evolution of the background entropy affects both the
evolution and the shape of the $L$--$T$ relation.  We already
emphasized the fact that the uncertainty in the evolution of the
$L$--$T$--$M$ relations reflects on the uncertainty in the derivation
of cosmological parameters from the cluster abundance evolution. The
$L$--$M$ relation is, in fact, the link between the cluster mass
function (predictable for a given cosmology with numerical or
analytical calculation) and the observed X--ray luminosity
distribution. The complexities due to the evolution in the luminosity
are only partially avoided when directly using the temperature.  In
fact, selection effects for flux limited samples add to the evolution
of the emission weighted temperatures (see Eke et al. 1998).

If the minimum background entropy $S_*$ is kept constant at every
epoch, the evolution of the $L$--$T$ relation is essentially frozen,
or mildly negative, even at redshifts as high as $z=1$, as already
shown in Figures \ref{fig11} and \ref{fig12}.  The evolution of $L$ at
fixed $T_{ew}$ is negative especially at small temperature.  This
global behaviour is in agreement with the claim for null evolution of
the $L$--$T$ at redshift $z\simeq 0.4$ (Mushotzky \& Scharf 1997).  A
non evolving $L$--$T$ relation, suggested also by the present data on
the luminosity function at high redshifts ($z>0.5$), would strengthen
a low density, eventually flat, universe (c.f. Borgani et al. 1999).

\begin{figure}
\centerline{\psfig{figure=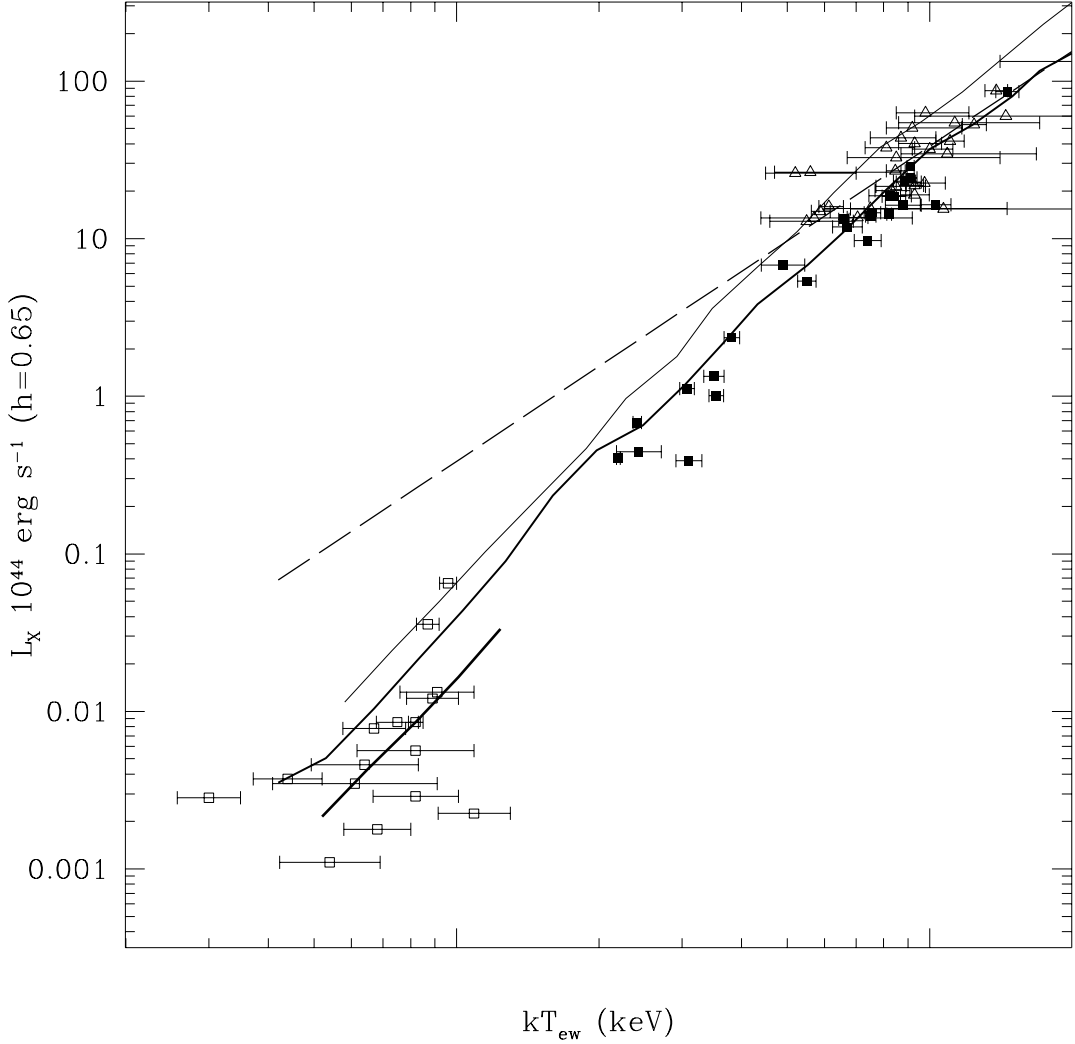,height=15truecm,width=15truecm}}
\caption{Relation between bolometric luminosity and emission
weighted temperature in $\Lambda$CDM assuming an evolving entropy
$K_*=0.8 (1+z)^{-1} \times 10^{34}$ erg cm$^2$ g$^{-5/3}$.  The thick
line for $k_BT\leq 1.5$ keV shows the $L$--$T$ relation at $z=0$
defined within the projected radius of $100 \, h^{-1} $ kpc as in
Ponman et al. (1996).  The dashed lines refer to the self--similar
case, while thick lines to $z_0=0$ and thin lines to $z_0=1$. Data as
in Figure \ref{fig11}.
\label{fig18}}
\end{figure}

\begin{figure}
\centerline{\psfig{figure=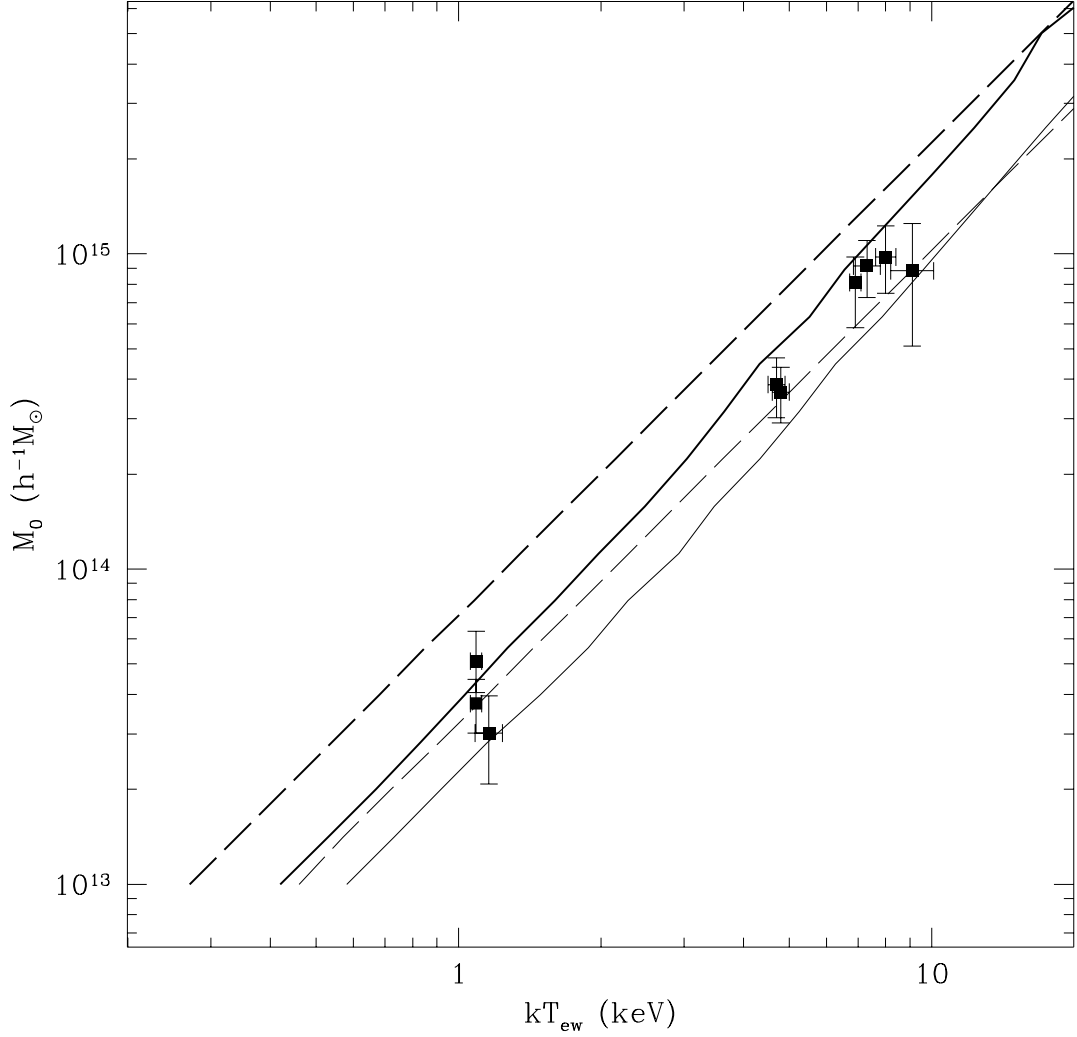,height=15truecm,width=15truecm}}
\caption{The $M$--$T$ relation in $\Lambda$CDM assuming an evolving
entropy $K_*=0.8 (1+z)^{-1} \times 10^{34}$ erg cm$^2$ g$^{-5/3}$.
The dashed lines refer to the self--similar case, while thick lines to
$z_0=0$ and thin lines to $z_0=1$. Data as in Figure \ref{fig13}.
\label{fig19}}
\end{figure}

We can investigate how the evolution of the $L$--$T$ is affected if
the background entropy evolves substantially with epoch.  In the
Figure \ref{fig18} and \ref{fig19} we assumed $K_{34}(z)=0.8 \,
(1+z)^{-1}$, which is an evolution that parallels the one observed in
the Ly$\alpha$ clouds.  In this scenario, objects observed at redshift
$z=1$ have accreted most of their baryonic mass when the entropy was
lower, and thus mostly in the shock regime. This allows the cooling to
start earlier, and be more efficient.  As a net result, the $L$--$T$
and the $K$--$T$ relations at $z=1$ are higher with respect to the
predictions of the constant $K_*$ scenario.  However the positive
evolution is about a factor of two, much less than the intrinsic
scatter, and very difficult to observe. Such a positive evolution is
too small to reconcile a critical universe with the observed high
redshift luminosity function.  As a further comment, we recall that
the large discrepancy between the average level of entropy seen in
Ly$\alpha$ clouds and that observed in the center of groups, implies
that the Ly$\alpha$ gas is not the same or the heating rate is much
steeper than this.  We therefore adopt this entropy evolution as
a reference case.

Assuming $K_{34}=0.8(1+z)^{-1}$,
gives a good fit to the whole temperature range without requiring
further dependence on the mass scale.  This is because the evolution
$(1+z)^{-1}$ introduces by itself such a dependence.  The core of
intermediate mass halos are assembled at $z\simeq 1$, for an effective
$K_{34}(z=1)=0.4$, while low mass objects build their cores at
redshifts $z\simeq 2-3$, for $K_{34}(z=3)=0.2$.  Also an evolution as
strong as $K_{34}=3(1+z)^{-2}$ provides a good fit to the data.

\section{DISCUSSION}

The main limitation of this model is clearly the adopted spherical symmetry
and also the assumptions of isotropic and continuous infall.  In the real
world, some of the baryons are accreted in the form of smaller clumps
and substructure, and flow along sheets and filaments.  The spherical
infall model used here does not include the effects of larger and
smaller scale perturbations. Moreover, there are missing ingredients
in the physics of baryons. We shall briefly discuss them in turn.

The presence of large--scale structure is not expected to affect
strongly the accretion rates and in general the statistical properties
of dark matter halos. In fact, the rates used in this work are derived
from the PS formalism, which proves accurate within few percent when
compared to N--body simulations that include large--scale structure
(see, e.g., Governato et al. 1998).  However, an effect of the
large--scale structure which is of interest here, could be the
eventual contribution to the initial entropy in the IGM due to shocks
occurring on large scales related to the formation of filaments.
Hierarchical gravitational processes do not break the
self--similarity, but the anisotropic collapse can produce widespread
shocks that raise the average entropy level in the IGM everywhere
without being associated with the formation of halos.  The baryons
that fall in the isotropic potential wells at the intersection of
sheets and filaments could be already heated by an amount which
depends on the power spectrum on large scales.  This can break the
self--similarity of the baryons, assuming that the large--scale
heating is effective almost uniformly in the IGM.

Focusing on smaller scales, the presence of substructure in the
infalling matter necessarily introduces some stochasticity in the
accreting processes.  The intrinsic scatter in the density and the
temperature of the accreted baryons translates into a scatter in the
observational quantities (see, e.g., CMT97). The presence of
substructures implies some gravitational energy is transferred to the
baryons before they are accreted into the main potential well and
shocked for the last time.  However, the gravitationally--produced
entropy on small scales is very different from the above mentioned
large scale production.  In fact, the mass distribution of satellite
halos scales self--similarly with the total mass of the final halo.
Thus the amount of entropy given to the baryons in substructures
scales with the final mass, and does not produce any break of
self--similarity.  This entropy contribution can be included in the
external entropy, $K_*$, without any distinguishing effect with
respect to the mass scale.

Another point related to the dark matter is the case of very massive
merger events, where a massive, disruptive event is defined by the
mass ratio of the merging halos being larger than about $0.3$ (see
Roettiger et al. 1998).  In these cases it is likely that the ICM is
strongly stirred, and, if the lookback time of the event is less than
$1$ Gyr, the ICM is not even in hydrodynamical equilibrium at the
epoch of observation.  Massive mergers can also create situations of
non--equilibrium ionization (see Ettori \& Fabian 1998; Takizawa
2000).  It is clear that the model cannot describe the population of
such disturbed clusters.  In the PS formalism, the fraction of objects
that are subject to large merger events is a sensitive function of
both the total virial mass $M_0$ and the observation epoch $z_0$.  We
calculate that the expected number of major mergers in the last Gyr is
between $0.1$ and $0.2$ in tCDM, and a factor of 2 lower in
$\Lambda$CDM, at $z_0=0$ (for a mass range between $10^{15}$ and
$10^{13}$ $h^{-1} M_\odot$).  However, such numbers grow to $1$--$0.5$
at $z_0=1$ in tCDM and $0.6-0.3$ in $\Lambda$CDM.  In this framework,
it is reasonable to expect that at $z\simeq 1$, a fraction between
$1/3$ and $1/2$ of the population of clusters has undergone a massive
merger event with a lookback time less than 1 Gyr.  This has to be
regarded as an intrinsic limitation to statistical analyses of the
population of high redshift clusters.

Other limitations come from the more complex physics of baryons.  An
important issue is that the entropy in the center may increase because
shocks propagate in the inner part of the halos due to infalling gas
along philaments (A. Klypin 2000, private communication).  We stress
however, that in order to survive the outer shock and propagate in the
very central part of the halo, the infalling baryons should be
compressed already.  The presence of an initial entropy level will
inhibit the formation of dense knots of gas at least on small scales,
and thus inner shocks are probably limited only to very massive
mergers.

Another important component, which is not included in the present
model, is the momentum gained from the heated gas, that can
push part of the baryons out of the halos without contributing to the
average heating.  This effect is very difficult to model a priori.
Its effect on the X--ray emission can be computed by including
semianalytical models of galaxy formation (see Menci \& Cavaliere
2000).

Finally, gravitational effects of the baryons on the dark matter
profile are neglected.  These can be important in the very center,
where the baryons can concentrate in the form of cooled gas and
contribute to density peaks which may affect the X--ray emission (see
Pearce et al. 2000, Lewis et al. 2000).

\section{CONCLUSIONS}

We have presented a detailed model to relate the X--ray properties of
diffuse baryons in clusters of galaxies to the entropy history of the
cosmic baryons, after including adiabatic compression, shock heating
and cooling.  Our aim is to build a useful tool to reconstruct the
entropy history of the universe from the observations of local and
distant clusters.  In particular, a major goal is to identify and
follow in time the processes that generate the entropy excess.  This
entropy excess is now probed by many observations and it is connected
with many scaling properties of X--ray halos.  Even if a given entropy
excess does not translate into a unique heating history, the
comparison of X--ray data with observations in other bands may allow
identification of the major heating sources.  Favoured candidates are
star formation processes and nuclear activity.  At present, however,
neither the epoch, nor the source of the related heating process have
been identified.

In this paper we have limited the investigation to a scenario in which
the excess entropy is present since very high $z$ and is uniform
throughout the IGM.  A case with an external entropy decreasing with
redshift, mimicking the rise of a population of heating sources, is
also presented.  In both the constant and time--evolving case, the
scaling properties of local clusters of galaxies are reproduced on a
large range of scales, with an appropriate choice of the free
parameter $K_*$.  The properties of distant X--ray halos are predicted
to be generally similar to properties of the local population, but
significative differences can be actually observed by the present--day
X--ray satellites, shedding light on the thermodynamics history of the
ICM.  We recall here the general results on density and temperature
profiles, together with the results on the evolution of the global
X--ray properties, especially luminosity and emission weighted
temperatures.

The bending of the $L$--$T$ relation with respect to the self--similar
case $L\propto T^2$, is due to the flatter profiles of the ICM going
from large mass to small mass halos.  Good fits are obtained for a
background entropy in the range $K_* = (0.2\pm 0.1) \times
10^{34}$ erg cm$^2$ g$^{-5/3}$ for $k_BT_{ew}<2$ keV, and $K_* = (0.4\pm
0.1) \times 10^{34}$ erg cm$^2$ g$^{-5/3}$ for $k_BT_{ew}>2$ keV.  This
scale dependence can be introduced by an evolution in the effective
value of $K_*$.  In particular, $K_{34} = 0.8 (1+z)^{-1}$ gives a good
fit over the whole range of observed temperatures.

The central regions of groups and clusters, which dominate the X--ray
emission, are formed during the initial stages of accretion.  In these
early phases, if a significant background entropy is present, the
accretion is adiabatic, and the gas is compressed in a flat, low
density profile with steep temperature gradients.  This is relevant
for the smallest halos, where the gravity does not overcome the
pressure support of the baryons for the majority of the subsequent
accretion of gas.  In clusters the infall velocities rapidly
become larger than the sound speed, and the shock regime takes over.
In the outer regions of clusters the entropy is entirely due to
gravitational processes, and the entropy profile is a featurless power
law approaching $K\propto r^{1.1}$.

This mechanism is particularly efficient if cooling is neglected.
However, it is known that the cooling is an important ingredient in
the history of the ICM.  The main effect is that the isentropic cores
expected in the constant entropy scenario, are partially erased by the
process of cooling.  Still, if $K_* > 0.1 \times 10^{34}$ erg cm$^2$
g$^{-5/3}$, the cooling processes are significantly suppressed and the
inner regions of the halos keep the imprint of the initial entropy
level.  Cooling processes appear again only in massive halos, where
the gravity dominates the energy of the system and the excess entropy
is no longer able to keep the gas at low density.  In the extreme case
of negligible $K_*$, it is worth noting that the cooling processes
alone would have a dramatic effect on both small and large mass halos.
In small mass halos ($10^{13} M_\odot$) most of the gas is expected to
cool and recombine, causing a central {\sl baryonic catastrophe}.

Other important characteristics are found in the temperature structure
especially of smaller halos.  Temperature gradients are commonly
expected both in clusters and in groups.  The polytropic index is
predicted to be $\gamma_p\simeq 0.9 -1.2$ in the region where the gas
is shock heated.  The polytropic index can be higher if the dark
matter profile is centrally peaked (e.g., with a power law with index
$\sim -1.4$, see Moore et al. 1998).  Another relevant observable (for
local halos) is the position of the final shock radius, which is
expected to be close to the virial one at large mass scale, while it
migrates to larger radii in small groups.  In the smallest halos, in
fact, the shock is vanishingly small.  As a function of epoch, for a
given object, the shock/accretion radius is initially quite distant
from the virial radius.  It is very close to the virial radius when
the mass accretion rate reaches its maximum and the shock regime is
well developed.  Eventually, the mass accretion rate decreases
(especially in the $\Lambda$CDM universe) and the shock radius relaxes
again to larger positions.  A consequence of the above picture is that
the ratio of the baryonic mass included in the virial radius to the
total mass, is always lower but still close to unity; it can be
significantly lower ($1/3$) only for small mass halos (corresponding
to emission--weighted temperatures of $0.3-1$ keV).

It is remarkable that the simple presence of an initial excess entropy
in the diffuse IGM can reproduce many of the scaling properties of the
observed X--ray halos, without the contribution of any internal
heating.  It is interesting to discuss the implications of this simple
scenario for the energetic budget and the past cosmic star formation
history. The minimum excess energy associated with an initial
background entropy $K_{34}$ is about
\begin{equation}
k_BT\simeq 0.1 \Big( {{K_{34}}\over{0.4}}\Big) \, \, {\rm keV},
\end{equation}
where the gas is assumed to be at the background density at the epoch
of the heating.  However, we can speculate on the energy budget when
the entropy excess is generated after the collapse, at much larger
densities (the internal scenario).

Following PCN, we can establish a relation between the epoch of
heating and the energy released.  Under the assumption that
the heating process can be described with a single epoch and a typical
overdensity, we have:
\begin{equation}
1+z_h = \Big( {{k_BT_h}\over{3.2 \times 10^{-2} K_{34}}}\Big)^{1/2}
\delta^{-1/3} \, ,
\end{equation}
where $k_BT_h$ is the average energy per particle released in the IGM by
non--gravitational processes.  If we adopt the conservative scenario
in which the gas is heated  at a typical virial density ($\delta
\simeq 200$), to have an entropy level in the range $K_* = (0.4\pm
0.2)\times 10^{34}$ erg cm$^2$ g$^{-5/3}$, we obtain:
\begin{equation}
1+z_h \simeq (1.5\pm 0.3)\Big( {{k_BT_h}\over{1 {\rm keV}}} \Big)^{1/2}\, .
\end{equation}
Thus, if we want heating at $z>1$ in order to avoid the overcooling
catastrophe, the energy budget must be larger than $1$ keV per
particle.  The above estimate would give even larger values after the
inclusion of cooling. In fact, if the gas is heated at high densities,
most of the extra energy is likely to be re--emitted soon, and this
would raise the energetic budget for a given final entropy
excess.  In this respect, the relation between the epoch of heating
and the energy released is strongly dependent on the physical process.
Of course, a scenario in which the extra entropy is provided by the
contributions of several different sources, active at different
epochs, is a likely possibility.  In this perspective, the measure of
metallicities as a function of the entropy of the baryons in different
systems, from Ly$\alpha$ clouds to rich clusters, may be useful in
determining whether the excess entropy is linked to star
formation processes.

The assumption of an initial excess entropy uniformly diffused in the
IGM, offers new perspectives in the approach to cluster formation, but
also galaxy formation.  Such an entropy background, once established,
may affect the star formation itself, since the cooling processes on
all scales are virtually inhibited.  This is the mechanism which is
expected to solve the {\sl cooling catastrophe} (see White \& Rees
1978, Blanchard, Valls Gabaud \& Mamon 1992; Prunet \& Blanchard 1999)
and in this view X--ray clusters and galaxy formation processes are
intimately related.  Current attempts to model {\sl ab initio} the
physics of the heating process, and then link the entropy history of
the cosmic baryons to galaxy formation, must include the well known
plethora of ingredients that has been already mentioned several times:
feedback from star formation processes and SNe explosions, radiative
and mechanical heating from active galactic nuclei, radiative heating
from hard X--ray background, gravitational heating on large scale
filaments (see Menci \& Cavaliere 2000; Valageas \& Silk 2000; Wu,
Fabian \& Nulsen 2000; Madau \& Efstathiou 1999; Cen \& Ostriker
1999).  Such different scenarios allow for different entropy histories
of the universe, determining both the spatial distribution and the
evolution of the entropy in the diffuse gas.

A promising strategy for the near future is to look directly for the
distribution of the entropy in the ICM (TSN00).  A direct consequence of
assuming a uniform entropy everywhere in the gas, is that the groups
are expected to be surrounded by large halos of low surface brightness
gas, spread out over radii much larger than the virial radius of dark
matter halos.  This low--density gas may have been missed by
observations with the ROSAT satellite, but can be detected by the XMM
satellite.  Its emission can enhance the total luminosity of the
groups by more than a factor of $3$, including the lowest energy bins
of $\simeq 0.1$ keV.  Another promising observational channel is the
absorption from metals in the gas seen against bright X or UV sources.
If the source of the background entropy is star formation, significant
pollution by metals is expected.

The model presented here is to be considered a useful tool to
interpret the observations of high redshift clusters, that will be
provided especially by the Chandra and XMM satellites.  Our aim is to
build a solid link between the thermodynamics of the diffuse cosmic
baryons and the emitting properties of X--ray halos, in order to be
able to {\sl reconstruct the entropy history of the universe, at high
and low redshifts, from spectral and imaging X--ray observations}.
This will help in understanding the source of the entropy excess and
the time evolution of the corresponding heating process.

We acknowledge discussions with S. Borgani, N. Menci, and P. Rosati.
We thank T.J. Ponman for discussions and for providing the data in
Figure \ref{fig17}. We thank R. Giacconi for discussions and
continuous encouragement. We thank the referee, Greg Bryan, for
detailed comments.  PT thanks ESO Garching for hospitality during the
completion of this work. This work has been supported by NASA grant
NAG 8-1133.

\appendix

\section{The infall velocity}

We find upper and lower limits for the infall velocity
of the accreted baryonic shells computed with Equation (\ref{v_i}).  The
last two terms of Equation (\ref{v_i}) depend on the densities of the
shell at accretion ($\rho_e$) and at turnaround
($\rho_{ta}$).  The values of the two densities are derived requiring
conservation of mass, and assuming that baryons and dark matter are
still not decoupled at turnaround.  In particular, the exact
value of $\rho_e$ depends on the validity of the Equation (\ref{mdot}),
which is based on the assumption that the total baryonic mass accreted
at every epoch is $f_B M_{v}$, where $f_B$ is the universal baryonic
fraction.  Such an assumption can be tested with numerical simulations,
and will not be discussed here.

We focus on the numerical uncertainty in the estimate
of the term $\Delta W$. The total work per unit baryonic mass done by
the gravitational potential on the baryonic shell is:
\begin{equation}
W\equiv \int_{R_{ta}}^{R_s} {{GM(<r)}\over{r^2}}dr\, ,
\end{equation}
where the integral is computed along the trajectory $r(t)$.  If the
total mass within the shell were constant, the solution would be
simply:
\begin{equation}
W = {{v_{ff}^2}\over 2}\equiv
\Big({{GM_{ta}}\over{R_S}}-{{GM_{ta}}\over{R_{ta}}}\Big)\, ,
\end{equation}
where $M_{ta}$ is the total mass initially contained in the
turn--around radius $R_{ta}$.  The free fall velocity $v_{ff}$ refers
to a test particle falling from turnaround to the shock radius,
experiencing a gravitational force always from the same amount of
matter.  However, the actual mass enclosed by a given baryonic shell
will depend on time.  We can write:
\begin{equation}
M(<r) = [f_B+(1-f_B)Y(t)]M_{ta}\, ,
\end{equation}
since the baryonic mass inside the shell is constant, but the amount
of dark matter can change by a time dependent factor $Y(t)$.  The
complete solution can now be formally written as:
\begin{equation}
W = {{v^2_{ff}}\over 2}+\Delta W = {{v^2_{ff}}\over 2} -
{{GM_{ta}}\over{r_{ta}}}\int^{R_s}_{R_{ta}}
{{(f_B+(1-f_B)Y(t)-1)}\over{r^2}}dr\, .
\label{wcorr}
\end{equation}
At this point we note that the amount of mass that is included in a
given baryonic shell along its trajectory is always larger than the
initial mass $M_{ta}$, since the collisionless shells of dark matter
fall faster than the baryonic shells, which, instead, are pressure
supported.  Here, we neglect the shell crossing and the detailed
behaviour in time, but we recall that we want the solution only at the
accretion radius, which usually occurs just inside the most external
caustic of the dark matter (see the self--similar model of
Bertschinger 1985).  Thus, we can safely assume that the total mass
can only grow inside the baryonic shell.  The mass excess $\Delta M/ M_{ta} =
(f_B+(1-f_B)Y(t)-1)$ can be described with a generic power law
dependence on the actual position $r(t)$ of the kind:
\begin{equation}
f_B+(1-f_B)Y(t)-1 =\Big( f_B+(1-f_B)Y_s-1\Big)
\Big[ \Big( 1 -\Big({{r(t)}\over{r_{ta}}}\Big)^\alpha \Big] 
\Big[ 1 - \Big( {{r_s}\over{r_{ta}}}\Big)^\alpha \Big]^{-1}
\label{ym1}
\end{equation}
where $\alpha > 0$, and $Y_s$ is the value at the accretion.  To
calculate $Y_s$ we must know the dark matter density profile at radii
larger than the virial radius.  We do not propose a specific model
here, instead we simply use the density profile as computed in
Bertschinger (1985) as a reasonable approximation at radii larger than
the virial one.  We can substitute Equation (\ref{ym1}) in Equation
(\ref{wcorr}) and integrate, obtaining an estimate of $W$ as a
function of $\alpha$.  To eliminate the dependence on $\alpha$, we
take the limit for small and large values of $\alpha$, to obtain the
upper and lower values for $\Delta W$:
\begin{equation}
\Delta W  =  [f_B+(1-f_B)(Y_s-1)]{{GM_{ta}}\over {r_{ta}}} 
\Big(
\Big[  {{1-x_s}\over {x_s}} + {{\ln(x_s)+1-x_s}\over
{\ln(x_s) x_s}} \Big] \pm  
 \Big[ {{1-x_s}\over {x_s}} - {{\ln(x_s)+1-x_s}\over
{\ln(x_s) x_s}} \Big]\Big)\, .
\label{dw}
\end{equation}
The last term in Equation (\ref{dw}) bounds the possible values for
$\Delta W$, assuming a monotonic increase of the total mass enclosed
by the infalling shell.  The upper and lower values turn out to be
between $10$ \% and $30$ \% during the mass history of a given halo,
and are plotted in Figure \ref{fig1} as dotted lines.  This reflects
our error in computing the infall velocities of the baryonic shells.
The uncertainty in the infall velocities does not strongly affect the
mass scale at which the adiabatic/shock transition occurs, since the
dependence of $v_i$ on the accreted mass is very steep when shocks
begin to appear.  This effect is related to the fast migration of the
accretion radius from $\simeq 2 R_v$ to $\simeq R_v$ (see Figure
\ref{fig3}).

\section{Cooling processes}

Here we discuss how to compute the effect of the radiative cooling on
each baryonic shell.  The treatment of the cooling is complex and
constitutes the largest uncertainty in modelling the X--ray emission
from clusters in present--day numerical simulations, since the
predicted luminosity of the central region can heavily depend on the
adopted resolution (see, e.g., Suginohara \& Ostriker 1998).

There is of course no difficulty in solving Equation (\ref{ent_evol}) as
long as $\tau_{cool} > \Delta t$ where $\Delta t$ is the time
resolution.  However, the time resolution needed increases
dramatically when the density increases and $\tau_{cool} \sim \Delta
t$, since the cooling is a runaway process.  Since our calculation is
based on a sequence of hydrostatic equilibria, and we do not want to
end up with an heavy computation effort, we propose to use a
reasonable time step (of the order of few tenths of Gyr) and solve {\sl
analytically} the energy Equation (\ref{ent_evol}) for each shell within
each time step.  To do this we first assume that the cooling
proceeds isobarically within $\Delta t$, and compute the new value of
the pressure after each step to take into account the new equilibrium
positions of each shell.

If the pressure is constant for each shell within $\Delta t$,  
the density can be expressed as a function of the adiabat $K$ only, to
give:
\begin{equation}
\rho = p^{1/\gamma} k^{-1/\gamma}\, ,
\end{equation}
where $\gamma =5/3$ and the variables are assumed to be normalized to
the shock values as usual.  The temperature is then:
\begin{equation}
t= k^{1/\gamma} p^{(\gamma -1)/\gamma} \, .
\end{equation}

Following Sutherland \& Dopita (1993), we define the normalized
cooling function $\Lambda_N\equiv \Lambda_{net} n_e n_i$, where $n_e$
is the electron number density and $n_i$ is the ion number density.
For an average metallicity $Z= 0.3 Z_\odot$ we can approximate $n_e \,
n_i = 0.704 (\rho_B/m_p)^2$.  The cooling time now can be expressed as a
function of the adiabat $K$ and the normalized cooling function
$\Lambda_N$:
\begin{equation}
\tau_{cool} \simeq 2.13 {{kT_{s0}m_p}\over{\mu \rho_{s0}}}
k^{2/\gamma} p^{(\gamma -2)/\gamma} \, \, \Lambda_N^{-1} \, ,
\label{taubeta}
\end{equation}  
where the subscripts ``$s0$'' refer to the value at the shock.  
To write an analytic expression, we approximate $\Lambda_N$ with a
polynomial form:
\begin{equation}
\Lambda_N = C_1 (kT)^\alpha + C_2 (kT)^\beta +C_3\, ,
\label{lambdan}
\end{equation}
where the exponents take the values $\alpha=-1.7$ and $\beta= 0.5$.
The constants depend on the assumed metallicity, and are chosen as in
Table \ref{tab2} in order to reproduce the cooling function of
Sutherland \& Dopita (1993) within few percent in the energy range
$k_BT>0.03$ keV.

Thus, using the canonical value $\gamma=5/3$, the cooling time can be
written as:
\begin{equation}
\tau_{cool} = C_\tau T_{s0} { { k^{6/5} p^{-1/5} }
\over
{ C_1\, T_{so}^{\alpha} \, k^{{3\over 5} \alpha} \, p^{{2\over 5} \alpha}
+ C_2\, T_{so}^{\beta} \, k^{{3\over 5} \beta} \, p^{{2\over 5} \beta}
+C_3}}\, .
\label{taubeta2}
\end{equation}
The constant $C_\tau$ factorizes out the terms that depend on the
shock condition, and can be written as:
\begin{equation}
C_\tau = 1.62\, \times 10^2 [f_B(1-f_{cool}-f_*)h^2(z)\, \delta_{so} g_{s0} 
]^{-1}\,  {\rm Gyr/keV},
\end{equation}
where $\delta_{s0}$ is the overdensity with respect to the critical
density at redshift $z$, $T_{s0}$ is the temperature at the shock and
$g_{s0}$ is the compression factor at the shock; $f_{cool}$ and $f_*$
are respectively the fraction of baryons cooled in the center and the
fraction of baryons locked into stars.

Equation (\ref{ent_evol}) can be recast in term of the adiabat $K$
only, and the final adiabat $k_f$ can be recovered implicitly from the
solution in the finite time step $\Delta t$ (expressed in Gyr):
\begin{equation}
\Delta t = C_\tau T_{s0} \int^{k_f}_{k_i}dk { { k^{1/5} p^{-1/5} }
\over
{ C_1\, T_{so}^{\alpha} \, k^{{3\over 5} \alpha} \, p^{{2\over 5} \alpha}
+ C_2\, T_{so}^{\beta} \, k^{{3\over 5} \beta} \, p^{{2\over 5} \beta}
+C_3}}\equiv F(k_i,k_f) \, 
\label{solution}
\end{equation}

In particular, the condition $F(k_i,0)< \Delta t$ determine if a shell
with initial entropy $k_i$ cools completely within $\Delta t$.  At
each epoch, the region comprised within the largest shell for which
$F(k_i,0)< \Delta t$ is included in the cooled fraction $f_{cool}$ and
excluded from the diffuse, emitting phase.

\section{Concentration parameters}

The concentration parameters of the dark matter profiles depend on
epoch and cosmology, as shown in the numerical works of Navarro et
al. (1997) or analytical models (see, e.g., Lokas 2000).  A general
trend is that lower mass halos are more centrally concentrated than
high mass halos by virtue of the higher redshift of formation.  For
the same reason, halos of the same virial mass, but observed at higher
redshifts, are less concentrated, since the difference in the average
density at the formation and at the observation is smaller with
respect to low redshift halos. The mass dependence of the
concentration parameter, however, can be well approximated with power
laws which change slightly as a function of epoch and cosmology.  In
this paper we used the following approximations:
\begin{eqnarray}
c = 8.5 \, M_{15}^{-0.086} \, \, \, \, \, {\rm \Lambda CDM,}\, \,  z=0\\
c = 5.4 \, M_{15}^{-0.070}\, \, \,\, \, {\rm \Lambda CDM,} \, \,  z=1\\
c = 5.5 \, M_{15}^{-0.070}\, \, \,\, \, {\rm tCDM,}\, \,  z=0\\
c = 4.4 \, M_{15}^{-0.046}\, \, \,\, \, {\rm tCDM,} \, \,  z=1
\end{eqnarray}
where $M_{15} \equiv M/(10^{15} h^{-1} M_\odot)$.

\newpage

\newpage

\clearpage

\newpage

\begin{deluxetable}{l l l l l l}
\footnotesize \tablenum{1} \tablecaption{Cosmological parameters.
\label{tab1}} \tablewidth{0 pt} 
\tablehead{ \colhead{Model} & $\Omega_M$ & $\Omega_\Lambda$ & $h$ &
$n$ & $\sigma_8$ } \startdata tCDM & $1.0$ & $0.0$ & $ 0.5$ & $0.8$ &
$ 0.55 $\nl $\Lambda$CDM &$0.3$ & $0.7$ & $ 0.7$ & $1.0$ & $1.1$ \nl
\enddata \tablecomments{$h$ is the Hubble constant in units of $100$
km/sec/Mpc, $\sigma_8$ is the amplitude of the fluctuations at the
scale of $8 h^{-1}$ Mpc, and $n$ is the primordial spectral index.  }
\end{deluxetable}

\begin{deluxetable}{l l l l l}
\footnotesize \tablenum{2} \tablecaption{Coefficients $A$ and $B$  for
the mass histories of the main progenitors.  
\label{tab2}} \tablewidth{0 pt} 
\tablehead{ \colhead{Model, $z_0$} & $10^{12}\, h^{-1} \, M_\odot$ &  $10^{13}
\, h^{-1} \, M_\odot$ & $10^{14}\, h^{-1} \,  M_\odot$ &$10^{15}\, h^{-1} \,  M_\odot$ } 
\startdata tCDM, $z_0=0$ & 2.30 - 0.60 & 2.51 - 0.87 & 2.56 - 1.40 &
2.46 - 2.32 \nl
tCDM, $z_0=1$ & 3.02 - 1.78 & 3.20 - 2.36 & 2.82 - 3.32 & 1.76 - 5.05 \nl
$\Lambda$CDM, $z_0=0$ & 1.50 - 0.12 & 1.86 - 0.18 & 2.14 - 0.46 & 2.38 - 0.94 \nl
$\Lambda$CDM, $z_0=1$  & 1.92 - 0.88 & 2.38 - 1.14 & 2.48 - 1.82 & 2.20 - 2.94 \nl 
\enddata 
\end{deluxetable}

\begin{deluxetable}{l l l l }
\footnotesize \tablenum{3} \tablecaption{Parameters for the cooling
function \ref{lambdan}.
\label{tab3}} \tablewidth{0 pt} 
\tablehead{ \colhead{Metallicity} & $C_1$ & $C_2$ & $C_3$ } 
\startdata 0 $Z_\odot$ & $1.19\times 10^{-4} $ & $6.3 \times 10^{-2}$ & $1.9
\times 10^{-2}$\nl 
$0.1$ $Z_\odot$  &$2.8\times 10^{-3}$ & $5.8 \times 10^{-2}$ & $4 \times 10^{-2}$ \nl
$0.3$ $Z_\odot$  &$8.6\times 10^{-3}$ & $5.8\times 10^{-2}$ & $6.3\times 10^{-2}$ 
\enddata 
\tablecomments{The units for $C_1$ are $10^{-22}$ erg cm$^3$ s$^{-1}$
keV$^{-\alpha}$; the units for $C_2$ are $10^{-22}$ erg cm$^3$ s$^{-1}$
keV$^{-\beta}$; the units for $C_3$ are $10^{-22}$ erg cm$^3$ s$^{-1}$. }
\end{deluxetable}


\newpage

\begin{references}

\reference{} Abadi, M.G., Bower, R.G., \& Navarro, J.F. 2000,
MNRAS, 314, 759
\reference{} Allen, S.W., \& Fabian, A.C. 1998, MNRAS, 297, 57
\reference{}Arnaud, M., \& Evrard, A.E. 1999, MNRAS, 305, 631
\reference{}Balbi, A., et al. 2000, ApJL submitted, astro-ph/0005124
\reference{}Balogh, M.L., Babul, A., \& Patton, D.R. 1999, MNRAS, 307, 463
\reference{}Bertschinger, E. 1985, ApJS, 58, 39
\reference{} Blanchard, A., Valls--Gabaud, D., \& Mamon, G.A. 1992,
A\&A, 264, 365
\reference{}Bond, J.R., Cole, S., Efstathiou, G., \& Kaiser, N. 1991, ApJ 379, 440
\reference{}Bondi, H. 1952, MNRAS, 112, 195
\reference{}Borgani, S., Rosati, P., Tozzi, P. \& Norman, C. 1999,
ApJ, 517, 40
\reference{} Bower, R.~G. 1991, MNRAS, 248, 332
\reference{} Bower, R.~G. 1997, MNRAS, 288, 355
\reference{} Bryan, G.L., \& Norman, M.L. 1998, ApJ, 495, 80
\reference{} Buote, D.A. 2000, MNRAS, 311, 176
\reference{} Burles, S., \& Tytler, D. 1999a, ApJ, 499, 699
\reference{} Burles, S., \& Tytler, D. 1999b, ApJ, 507, 732
\reference{} Cavaliere, A., \& Fusco Femiano, R. 1976, A\&A, 49, 137
\reference{}Cavaliere A., Menci N., Tozzi P. 1997, ApJ, 484, L21 (CMT97)
\reference{}Cavaliere A., Menci N., Tozzi P. 1998, ApJ, 501, 493
\reference{} Cavaliere, A., Menci, N., \& Tozzi, P. 1999, MNRAS, 308, 599
\reference{}Cen, R., \& Ostriker, J.~P. 1999, ApJ, 514, 1
\reference{} David, L.~P., Slyz, A., Jones, C., Forman, W., Vrtilek, S.~D., \& 
Arnaud, K.A.  1993, ApJ, 412, 479
\reference{} Eke, V. R., Cole, S., Frenk, C. S., \& Henry, J.~P. 1998,
MNRAS, 298, 114
\reference{} Ettori \& Fabian, A.C. 1998, MNRAS, 293, L33 
\reference{} Ettori \& Fabian, A.C. 1999, MNRAS, 305, 834 
\reference{} Evrard, A.~E. 1990, ApJ, 363, 349 
\reference{} Evrard, A.~E., \& Henry, J.~P. 1991, ApJ, 383, 95
\reference{} Fabian, A.~C., \& Nulsen, P.~E.~J. 1977, MNRAS, 180, 479
\reference{} Gioia, I., Henry, J~P., Maccacaro, T., Morris, S.~L., 
\& Stocke, J.T. 1990, 
ApJL, 357, 35
\reference{} Frenk, C.S., et al. 1999, ApJ, 525, 554 
\reference{} Fujita, Y., \& Takahara, F. 2000, ApJ, 536, 523
\reference{} Governato, F., Babul, A., Quinn, T., Tozzi, P., 
Baugh, C. M., Katz, N., Lake, G. 1999, MNRAS, 307, 949
\reference{} Gunn, J.~E., \& Gott, J.~R. 1972, ApJ, 176, 1 
\reference{}Henry J.~P. 1997, ApJ 489, L1
\reference{} Helsdon, S.F., \& Ponman, T.J. 2000, MNRAS, 315, 356
\reference{}Henry J.~P. 2000, ApJ, 534, 565
\reference{} Horner, D.J., Mushotzky, R.F., \& Scharf, C.A. 1999, ApJ,
520, 78
\reference{}Kaiser, N. 1986, MNRAS, 222, 323 
\reference{} Kaiser, N. 1991, ApJ, 383, 104 
\reference{}Knight, P.~A., \& Ponman, T.~J. 1997, MNRAS, 289, 955 
\reference{}Lacey, C., \& Cole, S. 1993, MNRAS, 262, 627 
\reference{}Lange, A.E., et al. 2000, sumbitted to Phys. Rev. D, 
astro-ph/0005004 
\reference{}Landau, L.~D., Lifshitz E.~M. 1959, {\sl Fluid
Mechanics} (London, Pergamon press), p. 329 
\reference{} Lewis, G.F., Babul, A., Katz, N., Quinn, T., Hernquist,
L., \& Weinberg, D. 2000, ApJ, 536, 623
\reference{} Lloyd--Davies, E.~J., Ponman, T.~J., \& Cannon,
D.B. 2000, MNRAS, 315, 689
\reference{} Lokas, E.L. 2000, MNRAS, 311, 423
\reference{}Loewenstein, M. 2000, ApJ, 532, 17
\reference{}Markevitch, M. 1998 ApJ, 504, 27
\reference{}Mathews, W.~G., \& Bregman, J.~N. 1978, ApJ, 224, 308
\reference{}Madau, P., \& Efstathiou, G. 1999, ApJL, 517, 9
\reference{}Menci, N.,  \& Cavaliere 2000, MNRAS, 311, 50
\reference{}Metzler, C.~A., \& Evrard, A.~E. 1994, ApJ, 437, 564 
\reference{}Mohr, J.J, \& Evrard, A.E. 1997, ApJ, 491, 38
\reference{}Mohr, J.J, Mathiesen, B., \& Evrard, A.E. 1999, ApJ, 517, 627
\reference{}Moore, B., Governato, F., Quinn, T., Stadel, J., \& Lake, 
G. 1998, ApJL, 499, 5
\reference{}Mushotzky, R.~F., \& Scharf, C.~A. 1997, ApJ, 482, 13
\reference{}Navarro, J.F., Frenk, C.S., \& White, S.D.M. 1997, \apj,
490, 493 
\reference{}Nevalainen, J., Markevitch, M., \& Forman, W. 2000, ApJ,
536, 73
\reference{} Pearce, F.R.,Thomas, P.A., Couchman, H.M.P. 1994, MNRAS,
268, 953
\reference{}Pearce, F.R., Thomas, P.A., Couchman, H.M.P., \&  Edge,
A.C. 2000, MNRAS, 317, 1029
\reference{} Pen, U. 1999, ApJL, 510, L1
\reference{} Ponman, T.~J., Bourner, P.~D.~J., Ebeling, H., \& B\"ohringer, H.
1996, MNRAS, 283, 690 
\reference{} Ponman, T.~J., Cannon, D.~B., \& Navarro, F.J. 1999,
Nature, 397, 135 (PCN) 
\reference{}Press, W.H., \& Schechter, P. 1974, \apj, 187, 425 
\reference{}Prunet, S., \& Blanchard, A. 1999, preprint, astro-ph/9909145
\reference{}Renzini, A. 1997, ApJ, 488, 35 
\reference{} Renzini, A. 1999, in Chemical Evolution from Zero to High
Redshift, Edited by Jeremy R. Walsh, Michael R. Rosa. Berlin: 
Springer-Verlag, 1999, p. 185
\reference{}Ricotti, M., Gnedin, N.~Y., \& Shull, J.~M. 2000, ApJ,
534, 41
\reference{} Riess, A.~G. et al. 1998, AJ, 116, 1009
\reference{} Roettiger, K., Burns, J., \& Loken, C. 1993, ApJ, 407, L53 
\reference{} Roettiger, K., Stone, J.~M., Mushotzky, R.~F. 1998, ApJ,
493, 62
\reference{} Rosati, P., Della Ceca, R., Norman,
C., Giacconi, R. 1998, ApJ 492, L21 
\reference{} Rosati, P., Borgani, S., Della Ceca, R., Stanford, A., 
Eisenhardt, P., \& Lidman, C. 2000, in Large Scale Structure
in the X-ray Universe, eds. Plionis, M. \& Georgantopoulos, I.,
Atlantisciences, Paris, France, p. 13
\reference{} Ryu, D., \& Kang, H. 1997, MNRAS, 284, 416
\reference{} Schaye, J., Theuns, T.,
Leonard, A., \& Efstathiou, G. 1999, MNRAS, 310, 57
\reference{} Schindler, S. 2000, in Clustering at High Redshift, ASP
Conference Series, Vol. 200. Edited by A. Mazure, O. Le Fèvre, and
V. Le Brun., p. 374
\reference{}Suginohara, T., \& Ostriker, J.P. 1998, ApJ, 507, 16
\reference{}Sutherland, R.~S., \& Dopita, M.~A. 1993 ApJS, 88, 235
\reference{}Takizawa, M., \& Mineshige, S. 1998, ApJ, 499, 82
\reference{}Takizawa, M. 2000, ApJ, 532, 183
\reference{}Tittley, E.R., \& Couchman, H.M.P. 2000, MNRAS, 315, 834
\reference{} Tozzi, P., Scharf, C., \& Norman, C. 2000, ApJ, 542, 106 (TSN00)
\reference{}Valageas, P., \& Silk, J. 2000, A\&A, 350, 752
\reference{} White, S.D.M., \& Rees, M. 1978, MNRAS, 183, 341
\reference{} White, S.D.M., Navarro, J.F., Frenk, C.S., 
\& Evrard, A.E. 1993, Nature, 366, 429
\reference{} Wu, K.~K.~S., Fabian, A.~C., \& Nulsen, P.~E.~J. 2000,
MNRAS, 318, 889
\reference{}Wu, K.~K.~S., Fabian, A.~C., \& Nulsen, P.~E.~J. 1999, 
MNRAS submitted, astro-ph/9910122

\end{references}
\end{document}